\documentclass[
 amsmath,amssymb,
 aps,
 twocolumn,
prx,
]{revtex4-1}

\usepackage{graphicx}
\usepackage{enumerate}
\usepackage[utf8]{inputenc}
\usepackage{xcolor}
\usepackage{hyperref}
\usepackage{floatrow}
\usepackage{float}

\usepackage{amsmath}
\usepackage{amssymb}
\usepackage{arydshln}

\usepackage{tikz}
\usetikzlibrary{positioning,arrows,decorations.markings}

\newcommand{\tjzm}{$t$-$J^z$~model}
\newcommand{\ket}[1]{\left\vert #1 \right\rangle}
\newcommand{\bra}[1]{\left\langle #1 \right\vert}
\newcommand{\sket}[1]{\vert #1 \rangle}
\newcommand{\sbra}[1]{\langle #1 \vert}
\newcommand{\abs}[1]{\left\vert #1 \right\vert}

\def\K{%
    \operatornamewithlimits{%
        \mathchoice{
            \vcenter{\hbox{\huge $\mathcal{K}$}}%
        }{
            \vcenter{\hbox{\Large $\mathcal{K}$}}%
        }{
            \mathrm{\mathcal{K}}%
        }{
            \mathrm{\mathcal{K}}%
        }
    }
}

\providecommand{\newoperator}[3]{%
	\newcommand*{#1}{\mathop{#2}#3}}
\providecommand{\renewoperator}[3]{%
	\renewcommand*{#1}{\mathop{#2}#3}}
	
\newoperator{\srot}%
	{\mathrm{rot}}{\nolimits}

\newoperator{\ent}%
	{\mathrm{ent}}{\nolimits}
	
\renewoperator{\Re}%
	{\mathrm{Re}}{\nolimits}
	
\renewoperator{\Im}%
	{\mathrm{Im}}{\nolimits}	
	
\makeatletter
\providecommand*{\diff}%
	{\@ifnextchar^{\DIfF}{\DIfF^{}}}
	
\def\DIfF^#1{%
	\mathop{\mathrm{\mathstrut d}}%
	\nolimits^{#1}\gobblespace}
	
\def\gobblespace{%
	\futurelet\diffarg\opspace}
	
\def\opspace{%
	\let\DiffSpace\!%
		\ifx\diffarg(%
			\let\DiffSpace\relax
		\else
			\ifx\diffarg[%
				\let\DiffSpace\relax
			\else
				\ifx\diffarg\{%
					\let\DiffSpace\relax
				\fi\fi\fi\DiffSpace}

\newcommand{\mean}[1]{\langle#1\rangle}

\usepackage{bm}	
\renewcommand{\vec}[1]{\bm{#1}}

\begin{document}

\title{Anomalous Eigenstates of a Doped Hole in the Ising Antiferromagnet}



\author{Piotr Wrzosek$^1$}
\author{Krzysztof Wohlfeld$^1$}
\author{Eugene Demler$^{2}$}
\author{Annabelle Bohrdt$^{3,4}$}
\author{Fabian Grusdt$^{3,4}$}

\affiliation{%
$^1$Institute of Theoretical Physics, Faculty of Physics, University of Warsaw, Pasteura 5, PL-02093 Warsaw, Poland
}%

\affiliation{%
$^2$Institute for Theoretical Physics, ETH Zurich, Zurich, Switzerland
}%

\affiliation{%
$^3$Department of Physics and Arnold Sommerfeld Center for Theoretical Physics (ASC),
Ludwig-Maximilians-Universit\"at M\"unchen, Theresienstr. 37, M\"unchen D-80333, Germany
}%

\affiliation{%
$^4$Munich Center for Quantum Science and Technology (MCQST), Schellingstr. 4, D-80799 M\"unchen, Germany
}%

\date{\today}

\begin{abstract}
The problem of a mobile hole doped into an antiferromagnet Mott insulator is believed to underly the rich physics of several paradigmatic strongly correlated electron systems, ranging from heavy fermions to high-Tc superconductivity. Arguably the simplest incarnation of this problem corresponds to a doped Ising antiferromagnet, a problem widely considered essentially solved since almost 60 years by a popular yet approximate mapping to a single-particle problem on the Bethe lattice. Here we show that, despite its deceptive simplicity, the local spectrum of a single hole in a classical Ising-N\'eel state contains a series of anomalous, long-lived states that go beyond the well-known ladder-like spectrum with excited energies spaced as $J^{2/3} t^{1/3}$. The anomalous states we find through exact diagonalization and within the self-avoiding path approximation have excitation energies scaling approximately linear with $J$ and lead to a series of avoided crossings with the more pronounced ladder spectrum. By also computing different local, rotational spectra we explain the origin of the anomalous states as rooted in an approximate emergent local $C_3$ symmetry of the problem. From their direct spectral signatures we further conclude that these states lead to anomalously slow thermalization behavior -- hence representing a new type of quantum many-body scar state, potentially related to many-body scars predicted in  lattice gauge theories. 
\end{abstract}

\pacs{Valid PACS appear here}
\maketitle


\section{\label{sec:intro}Introduction}

Understanding the motion of a single hole in a two-dimensional (2D) Ising antiferromagnet, as governed by the $t$--$J^z$ model, can be seen as a canonical strongly correlated electron problem. Indeed, it was first studied by Bulaevskii, Nagaev and Khomskii already in 1968~\cite{Bul68}. In~\cite{Bul68,Bri70,Kan89} it was proposed that such a problem can be mapped onto the motion of a single particle in an external discrete linear (string) potential known from quark confinement. This leads to the discrete energy levels that scale as $J^{2/3} t^{1/3}$ with the model parameters (tunneling $t$ and Ising coupling $J$). The resulting spectral function consists of a set of well-defined Dirac delta peaks, each split from another by a distance $\propto J^{2/3} t^{1/3}$, and with their spectral weights scaling as $Z\sim J/t$ when $J \ll t$~\cite{Kan89,Bohrdt2020}. Subsequent theoretical studies have confirmed that the dominant features of the one-hole spectrum in the Ising antiferromagnet follow the ladder spectrum described above, see e.g.~\cite{Che99,Wrzosek2021}.

The persistent interest in the above problem derives from its close relation to the issue of high-Tc superconductivity that was first discovered in the cuprates in the mid-1980s. For this reason, it has been extensively studied in the late 1980s and early 1990s, when many of its key features were derived~\cite{Tru88,Kan89,Sachdev1989,Simons1990,Beran1996,Che99}. Among them was the realization by Trugman~\cite{Tru88} that certain hole paths involving loops can lead to a free propagation of the hole through the N\'eel state, instead of being fully localized as originally proposed by Bulaevskii {\it et al}.~\cite{Bul68}. However, the quantitative effect of such Trugman loops is very small in the Ising antiferromagnet~\cite{Tru88,Gru18,Wrzosek2021} such that it is sufficient to consider the {\it local} spectral function which is not broadened significantly due to Trugman loops. 

A study by Simons and Gunn~\cite{Simons1990} addressed the states beyond the ladder spectrum. They revealed a rich internal structure of the one-hole states, which was later rediscovered and associated with ro-vibrational excitations of the string created by the doped hole~\cite{Gru18}. Moreover, a close relation of these internal string states to confined mesonic excitations~\cite{Greensite2003} of gauge theories was established~\cite{Beran1996,Gru18}. These internal, ro-vibrational states will also play a central role in our present paper.

\begin{figure*}[t!]
\begin{center}
	\includegraphics[width=\columnwidth]
	{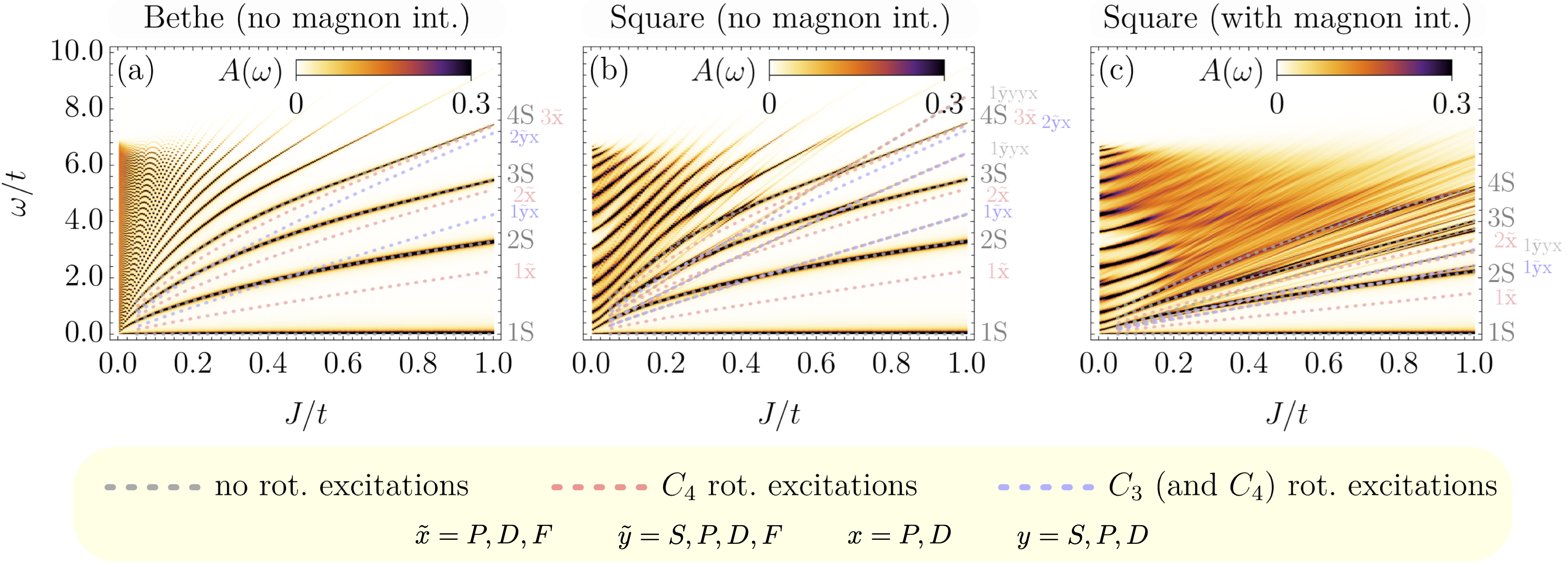}
	\caption{We compute the local spectral function $A(\omega)$ of a single hole introduced to the \tjzm{} for different strengths of the spin interaction $J/t \in [0, 1]$ of the \tjzm{}. Energies $\omega$ are measured relative to the ground state (lowest-energy peak). Calculations were performed using the self-avoiding walks approximation ({\it cf}. \cite{Wrzosek2021} and text for further details). (a) On the Bethe lattice, where an exact analytical solution is possible, $A(\omega)$ consists only of the ladder spectrum. The corresponding peaks ($nS$) can be labeled by their vibrational quantum number $n=1,2,...$ and have no angular momentum ($S$-wave states). On the square lattice, but neglecting magnon-magnon interactions, (b), a new set of anomalous states is found in this work, which lead to avoided level crossings with the more pronounced ladder states. We show in this article that this new set of states $nSx$ can be labeled by an approximate $C_3$ angular momentum $x$. Importantly, both the ladder states and the new anomalous peaks remain clearly visible, on top of an emerging incoherent background, when magnon-magnon interactions on the square lattice are included (c). Dashed lines on top of each spectrum represent the locations of the most pronounced low lying vibrational and rotational modes of the hole identified throughout the paper, using rotational spectral functions which can also resolve states with non-zero $C_4$ lattice angular momentum (e.g. the series $n\tilde{x}$ with $\tilde{x}=P,D,F$). Under idealized conditions in a), rotational excitations have zero spectral weight and dashed lines only serve as a reference. In b), c), dashed lines help identify avoided crossings between different ro-vibrational states.}
	\label{fig:3panel}
	\end{center}
\end{figure*}

Here we study the dynamical properties of a single hole in a N\'eel state, and demonstrate that a series of anomalous eigenstates exist in the spectrum which closely resemble quantum many-body scars discovered in a range of system in recent years~\cite{Bernien2017,Turner2018,Surace2020,Serbyn2021}. Such quantum scars constitute a subset of non-typical eigenstates which do not obey the eigenstate thermalization hypothesis (ETH) and co-exist with a continuum of typical states that fall into the ETH paradigm. In addition to narrow peaks in the spectral function -- on which we will focus here, see Fig.~\ref{fig:3panel} -- scarred states lead to pronounced oscillations following non-equilibrium quenches~\cite{Bernien2017}. Interestingly, such oscillations have been predicted in real-time, real-space dynamics of a single hole released in an antiferromagnetic background~\cite{Golez2014,Gru18,Bohrdt2020NJP,Nielsen2022}, and have been argued to be related to the specific structure of the ladder spectrum described above. These can be viewed as quantum scars, associated with the vibrational excitations of string states. As we will show in this paper by a detailed investigation of the one-hole spectrum, an additional series of scar states exist in the $t-J^z$ model which are associated to rotational excitations of the strings. 

Despite the breadth and depth of the early studies of the one-hole problem in the Ising background, the detailed structure of the corresponding spectra has remained surprisingly poorly understood. With the advance of new numerical techniques and analytical approaches, much more quantitative studies can be performed in larger systems and important new qualitative insights can be obtained. Among them was the recent clarification that the ubiquitous ladder spectrum of a single hole co-exists with a broad, incoherent background of many-body states. The origin of this featureless part of the spectrum was explained by some of us to be rooted in magnon-magnon interactions~\cite{Wrzosek2021} -- or equivalently, using string language, in self-interactions among strings~\cite{Bermes2024}. This can be readily seen by comparing panels (b) and (c) in Fig.~\ref{fig:3panel}, where the incoherent spectrum emerges as magnon-magnon interactions are included. 

The main result of the present paper is the identification of a second series of anomalous, scar states in the one-hole spectrum, in addition to the states from the ladder spectrum. These new states are absent in the one-hole spectrum on the Bethe lattice, see Fig.~\ref{fig:3panel} (a) where only discrete ladder-type states are found. (As we show below, this is true even after the inclusion of magnon-magnon interactions). We report direct signatures of the new series of anomalous states in the local spectrum calculated on the square lattice, in the absence of magnon-magnon interactions. As shown in Fig.~\ref{fig:3panel} (b), additional spectral lines, accompanied by avoided level crossings with the standard ladder states, appear in this situation. Their excitation energies scale approximately linear with the magnon energy $J$, which indicates their relation to the rotational string states predicted earlier under {\it idealized} conditions~\cite{Simons1990,Gru18}. We further demonstrate that the additional anomalous states remain visible as robust peaks in the local spectrum of the full $t-J^z$ model which includes magnon-magnon interactions, see Fig.~\ref{fig:3panel} (c), supporting their quantum scar nature.

Through a detailed analysis of the rotational one-hole spectra below in the paper, we moreover uncover the microscopic nature of the newly found peaks in the spectrum of the $t-J^z$ model. This way we identify an approximately, but not fully conserved discrete $C_3$ angular momentum of the string states which is responsible for the formation of these exotic quantum many-body scar states. This also allows us to label all anomalous states in the spectrum by their ro-vibrational quantum numbers~\cite{Gru18}, as explained in the caption of Fig.~\ref{fig:3panel}. Finally, in the different rotational spectra we analyze, we also find further anomalous states with an exactly conserved $C_4$ angular momentum in the local one-hole spectrum. This angular momentum corresponds to the discrete $C_4$ symmetry of the square lattice. The locations of these further states are indicated by the additional dotted lines in Fig.~\ref{fig:3panel} (types labeled $nx...$ with $x=P,D,F$), which would become visible when the rotational symmetry of the $t-J^z$ model is weakly broken.

The results obtained in this paper not only advance our fundamental understanding of the paradigmatic $t-J^z$ model of a doped antiferromagnet; they are also of immediate experimental relevance, e.g. for quantum simulation experiments directly realizing clean $t-J^z$ Hamiltonians. Such simulations have originally been proposed more than a decade ago~\cite{Gorshkov2011tJ}, and have recently been realized using ultracold molecules~\cite{Carroll2024} and using an alternative scheme~\cite{Homeier2024} with Rydberg atoms in optical tweezers~\cite{Qiao2025}. The internal states of the molecules and atoms used in these studies can be utilized to perform out-coupling spectroscopy~\cite{Dao2007,Kollath2007,Bohrdt2018,Brown2019} to directly measure the local one-hole spectrum we study; multi-photon extensions to the more complex rotational spectra discussed below have also been suggested~\cite{Bohrdt2021PRL}. Finally, out-of equilibrium dynamics of individual holes in an antiferromagnetic background have been realized in neutral atom quantum simulation platforms~\cite{Ji2021,Qiao2025}, which offer an alternative way to probe the anomalous states discussed in this paper.

\section{\label{sec:model_methods}Model and Methods}
The Hamiltonian $\hat{H}$ of the \tjzm{} we study is defined in the usual manner~\cite{Bul68, Kan89,
Bri70, Tru88, Mar91,
Sta96,
Che99, Bie19, Wrzosek2021}:
\begin{align}
\hat{H} &= \hat{H}_t + \hat{H}_J, \label{eq:hamiltonian_1}
\end{align}
where
\begin{align}
	\hat{H}_t &= -t\sum_{\mean{\vec{i}\vec{j}}\sigma} \tilde{c}_{\vec{i}\sigma}^\dag \tilde{c}_{\vec{j}\sigma}
	+ \mathrm{H.c.}, \\
	\hat{H}_J &= J\sum_{\mean{\vec{i}\vec{j}}} \left(\hat{S}_{\vec{i}}^z \hat{S}_{\vec{j}}^z - \frac{1}{4}\tilde{n}_{\vec{i}} \tilde{n}_{\vec{j}}\right).
\end{align}
Here electron double occupancies are projected out, i.e. $\tilde{c}_{\vec{i}\sigma}^\dag = \hat{c}_{\vec{i}\sigma}^\dag (1 - \hat{c}_{\vec{i}\bar{\sigma}}^\dag \hat{c}_{\vec{i}\bar{\sigma}})$, and $\tilde{n}_{\vec{i}} 
= \sum_{\sigma} \tilde{n}_{\vec{i} \sigma} =
\sum_{\sigma} \tilde{c}_{\vec{i}\sigma}^\dag \tilde{c}_{\vec{i}\sigma}$
and the $\hat{S}^z$ spin operator is defined in terms of the electron operators as
$\hat{S}^z_{\vec{i}} = ( \tilde{n}_{\vec{i} \uparrow} - \tilde{n}_{\vec{i} \downarrow}) /2 $; above we sum over $\sigma=\uparrow, \downarrow$.
While $\ket{{\rm N}}$ is the undoped N\'eel AF ground state of the \tjzm{} model, that is a well-known product state, the eigenstates of this model with a single
hole are complex many-body states -- as discussed, {\it inter alia}, in this paper.

\begin{figure}[t!]
	\begin{center}
		\includegraphics[width=\columnwidth]
		{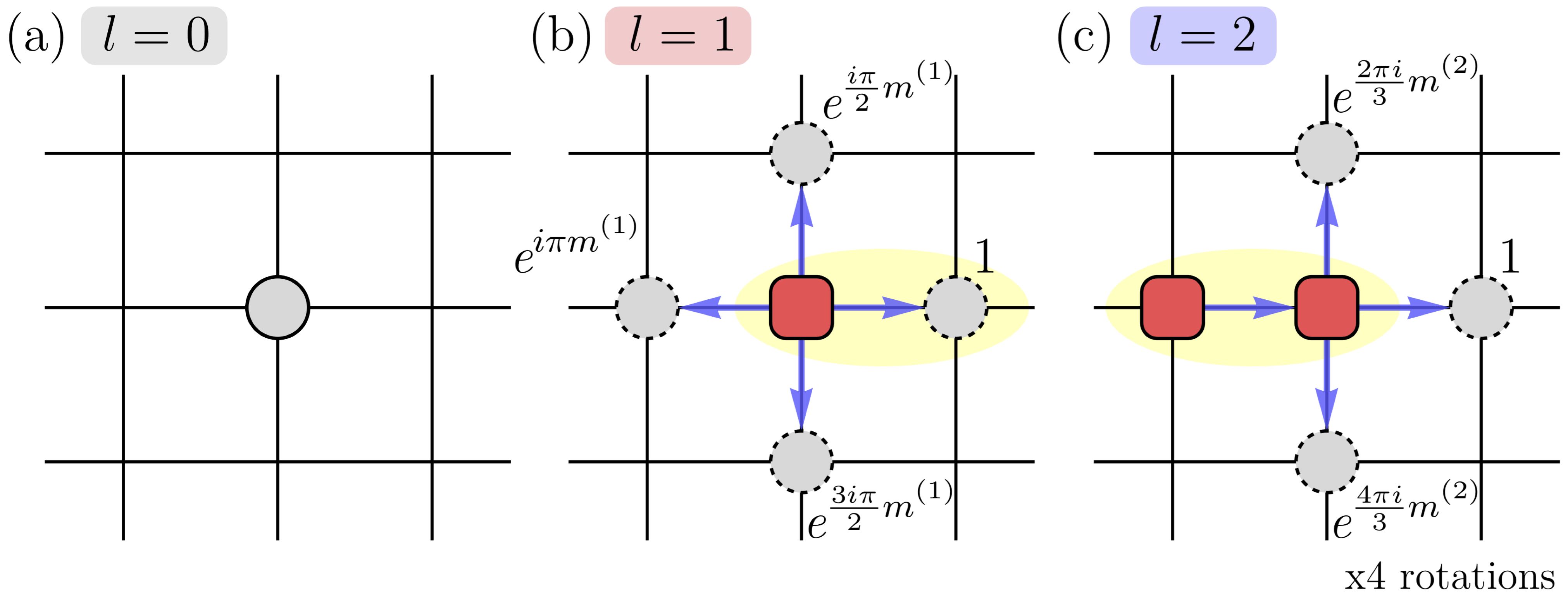}
		\caption{Cartoon picture of an initial state with (a) $l = 0$, (b) first-generation $l = 1$, and (c) second generation $l = 2$ rotational excitations. Gray circle represents a hole while red square (with rounded corners) stands for a magnon (string). The hole delocalized on multiple sites is denoted with dashed edges -- it also means a sum over multiple configurations of the hole and magnons on the lattice: (b) 4 configurations and (c) 12 configurations. Yellow area covers the corresponding sites in panels (b) and (c). In panel (c) only one of four rotations around the origin is shown. Varying $m^{(l)}$ yields different rotational states.
        }\label{fig:rot_cartoon}
	\end{center}
\end{figure}

The main object of interest in this study are rotational spectral function $ A_{M_l}(\omega)$ of different generations $l$:
\begin{equation}\label{eq:ag}
    A_{M_l}(\omega) =  -\frac{1}{\pi} {\rm Im}  G_{M_l}(\omega),
\end{equation}
where the 
\emph{local} rotational Greens function at site $\vec{i}$  reads (we set $\hbar=1$ throughout this paper)
\begin{equation}\label{eq:greens_function}
    G_{M_l}(\omega) = 
    \bra{{\rm N}}\hat{R}^{\dag}_{\sigma,M_l}(\vec{i}) 
        \frac{1}{\omega - \hat{H} + E_0}
    \hat{R}_{\sigma,M_l}(\vec{i}) \ket{{\rm N}}.
\end{equation}
Clearly, in this notation, the precise form of the rotational Green's function 
depends on the definition 
of the operator $\hat{R}$:

In general, the operator $\hat{R}_{\sigma,M_l}(\vec{i} \equiv \vec{0})$ annihilates an electron with spin $\sigma$ (i.e. creates a hole)
at site $\vec{i} = \vec{0}$ (the `origin' of the lattice), see Fig.~\ref{fig:rot_cartoon}(a). Then the hole is propagated $l$ times without returning through all possible paths starting at site $\vec{i} = \vec{0}$. Each time the hole is propagated, $n=1,2,...,l$, to the next site, it acquires a phase proportional to $M_l(n) = m^{(n)}$; here $m^{(1)} = 0,1,2,3$ is a $C_4$ (lattice) angular momentum quantum number, see Fig.~\ref{fig:rot_cartoon}(b) and $m^{(n>1)} =0,1,2$ denotes $C_3$ (lattice) angular momentum quantum numbers~\cite{Simons1990,Grusdt2018PRX}, e.g. for $l=2$ see Fig.~\ref{fig:rot_cartoon}(c); we will refer to $l=0,1,2,...$ as the \emph{generation} of rotational excitations. 
Intuitively, we can understand the action of such rotational operator as if the injected hole
in the AF lattice `initially' (i.e. by $l$ first hops) can only excite a particular superposition of magnons -- that is defined by a given value of the lattice angular quantum number. While the consecutive propagation of the hole is in principle not constrained to exciting a particular superposition of magnons, this initial `injection' of the lattice angular momentum has a lasting effect on the way the hole is allowed to propagate. 

In this work we study cases of $l = 0, 1, 2$. Note that the $l = 0$ case is special, for then the hole is {\it not} propagated to the neighboring sites by the operator $\hat{R}$ and  $M_l= \varnothing $. Thus, this case corresponds to the standard local Greens function of a single hole in the~\tjzm{}, which is studied e.g. in \cite{Kan89, Mar91, Che99, Bie19, Wrzosek2021}
and is {\it a priori} accessible in typical photoemission experiments. In principle higher-generation rotational operators,
i.e. those with $l >2$, can
be recursively defined following from case $l=2$ below. However, calculating such higher-order rotational spectra is a complex task that does not seem to bring any novel physical insight w.r.t. first second generation ($l=1,2$) cases studied here and therefore these are not considered below.

\begin{figure}[t!]
	\begin{center}
		\includegraphics[width=\columnwidth]
		{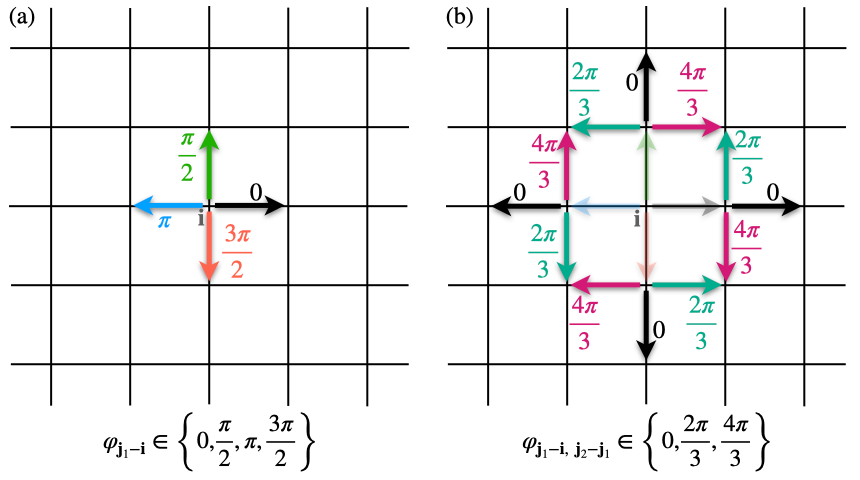}
		\caption{Cartoon picture of a square lattice with denoted direction dependent phase factors $\varphi$ of the Fourier transform used in the definition of the operator $\hat{R}_{\sigma,M_l}(\vec{i})$. The hole is introduced at site $\vec{i}$. We consider cases where the hole can be initially propagated up to $l=2$ times: $\mathrm{(a)}$ for first-generation excitations $l=1$ the hole acquires a phase $\varphi_{\vec{j_1}-\vec{i}} \in \{0, \frac{\pi}{2}, \pi, \frac{3\pi}{2}\}$; the values wind anticlockwise, where $\varphi_{\vec{j_1}-\vec{i}} = 0$ corresponds to the positive $x$ direction from site $\vec{i}$; $\mathrm{(b)}$ for second-generation excitations $l=2$ the hole acquires a phase $\varphi_{\vec{j_1}-\vec{i}} \in \{0, \frac{\pi}{2}, \pi, \frac{3\pi}{2}\}$ and then phase $\varphi_{\vec{j_1}-\vec{i}, \vec{j_2}-\vec{j_1}} \in \{0, \frac{2\pi}{3}, \frac{4\pi}{3}\}$. Since we consider forward propagation of the hole (no returns), there are exactly three possible directions for the propagation of the hole after the first hop. Note that the choice of phases $\varphi_{\vec{j_1}-\vec{i}, \vec{j_2}-\vec{j_1}}$ preserves the $C_4$ symmetry around site $\vec{i}$. 
        }\label{fig:varphi}
	\end{center}
\end{figure}

More precisely, the three distinct rotational Green's functions considered in this paper are defined through
the operator $\hat{R}$ 
that can create:
\begin{enumerate}
    \item A single hole without rotational excitations (zeroth generation), i.e.
\begin{equation}\label{eq:R_0}
    \hat{R}_{\sigma, \varnothing } (\vec{i}) \equiv \tilde{c}_{\vec{i}\sigma},
\end{equation}
where in this particular case set $M_l = \varnothing$;
this process is illustrated in Fig.~\ref{fig:rot_cartoon}(a) and corresponds to a `standard' single-hole removal spectral function, as already mentioned above.
\item A single hole with a first-generation rotational excitation, i.e.
\begin{align}\label{eq:R_1}
&\hat{R}_{\sigma,m^{(1)}}(\vec{i}) =  \mathcal{N}_1\sum_{\vec{j}: \langle \vec{j}, \vec{i} \rangle} \exp \left (i m^{(1)}  \varphi_{\vec{j}-\vec{i}} \right) \nonumber 
\\
& \times \left( \sum_\tau \tilde{c}_{\vec{i} \tau}^\dag \tilde{c}_{\vec{j}\tau} \right) \tilde{c}_{\vec{i}\sigma},
\end{align}
with the $C_4$ angular momentum quantum number $m^{(1)}=0,1,2,3$, $\mathcal{N}_1=1/2$ a normalization factor and $\varphi_{\vec{j}-\vec{i}}$ denoting the direction dependent phase factor of a Fourier transformation from real space to rotational basis, see Fig.~\ref{fig:varphi}(a);
this process is illustrated in
Fig.~\ref{fig:rot_cartoon}(b), i.e. in this case the hole  
creates string segments of length $l=1$ `directly' after it is created.
\item A single hole with second-generation rotational excitations, i.e.
\begin{align}\label{eq:R_2}
    &\hat{R}_{\sigma,m^{(1)},m^{(2)}}(\vec{i}) =
   \mathcal{N}_1 \mathcal{N}_2 \sum_{\vec{j}_1: \langle \vec{j}_1, \vec{i} \rangle} \sum_{\substack{\vec{j_2}: \langle \vec{j}_2, \vec{j}_1 \rangle\\\vec{j_2} \neq \vec{i}}} \nonumber \\
   &\times \exp \left( i m^{(2)} \varphi_{\vec{j}_1-\vec{i},\vec{j}_2-\vec{j}_1} \right) \exp \left( i m^{(1)} \varphi_{\vec{j}_1-\vec{i}}  \right) 
   \nonumber
   \\
   &\times \left(  \sum_{\tau_2}  \tilde{c}_{\vec{j}_1 \tau_2}^\dag \tilde{c}_{\vec{j}_2\tau_2} \right) \left( \sum_{\tau_1} \tilde{c}_{\vec{i} \tau_1}^\dag \tilde{c}_{\vec{j}_1\tau_1} \right) \tilde{c}_{\vec{i}\sigma},
\end{align}
with the $C_4$ angular momentum quantum number $m^{(1)}$ defined as above, the $C_3$ angular momentum quantum numbers $m^{(2)}=0,1,2$,
$\mathcal{N}_{1} = 1/2$,
$\mathcal{N}_{2} = 1/\sqrt{3}$, and the direction dependent phase factor $\varphi_{\vec{j}_1-\vec{i} \vec{j}_2-\vec{j}_1}$, see ~Fig.~\ref{fig:varphi}(b); 
this process is illustrated in Fig.~\ref{fig:rot_cartoon}(c) and corresponds to a hole 
creating string segments of length $l=2$ directly after the hole is created.
\end{enumerate}

\begin{figure}[t!]
	\begin{center}
		\includegraphics[width=\columnwidth]
		{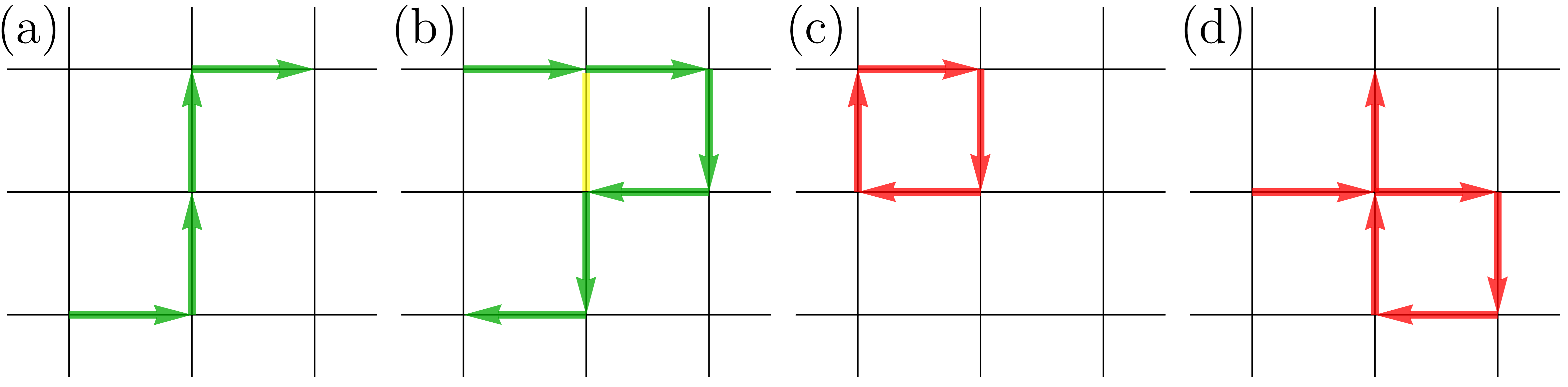}
		\caption{Different types of hole paths in the AF square lattice. Panels (a) and (b) show examples of non-crossing paths -- these type of paths are included in the SAW approximation~\cite{Wrzosek2021}. Note that this also includes a tangential path, where the bond at which the path becomes tangential is highlighted in yellow. Panels (c) and (d) show examples of paths with loops~\cite{Tru88} -- those are excluded from the considerations in our calculations.
        }\label{fig:saw_paths}
	\end{center}
\end{figure}

To evaluate the Greens function defined in Eq.~\eqref{eq:greens_function}, one can reliably use the so-called self-avoiding walks approximation (SAW)~\cite{Wrzosek2021}. Within SAW 
$\hat{H}_J$ is taken into account exactly, while the electronic hopping $\hat{H}_t$ is subject to a relatively minor approximation~\cite{Wrzosek2021}. In the half-filled ground state $\ket{{\rm N}}$ all the electrons are immobile due to the no double occupancies constraint in $\hat{H}_t$. Next, $\hat{R}_{\sigma,M_l}(\vec{i})$ removes single electron from $\ket{{\rm N}}$, i.e. single hole is created. The SAW simply states that we include all the possible hole hopping processes such that the created hole never crosses its own path. The examples of possible and excluded paths are shown in Fig.~\ref{fig:saw_paths}. Two self-avoiding walks can be observed in Fig.~\ref{fig:saw_paths}(a-b), where in Fig.~\ref{fig:saw_paths}(b) the path is also tangential -- spins of two electrons that belong to the path interact along the bond that does not belong to the path (bond highlighted in yellow). It is worth mentioning that taking this `off-path' interaction into account is equivalent to including non-crossing diagrams in the expansion of the Greens function 
-- however, this goes beyond the standard self-consistent Born approximation (SCBA)~\cite{Mar91} since the magnon-magnon interactions {\it can} be included in SAW. On the other hand, the paths that are crossing [i.e. paths that contain loops, see Fig.~\ref{fig:saw_paths}(c-d)] are excluded. Note that the latter are the so-called Trugman loops~\cite{Tru88} that were recently shown to have quite negligible weight in the single-hole problem of the 2D
$t$--$J^z$ model at the typically considered value of $J \ge 0.4t$~\cite{Wrzosek2021} (which basically stays in line with the considerations of \cite{Bar89, Che99, Bie19}).


The original formulation of the SAW approximation in Ref.~\cite{Wrzosek2021} allows for calculation of the diagonal terms of the Greens function -- these are the terms where the position of the hole is {\it the same} in the state on the left- and the right-hand side of the operator $(\omega - \hat{H} + E_0)^{-1}$ in Eq.~\eqref{eq:greens_function}. 
Note that once $l>0$ in the rotational operator $\hat{R}_{\sigma,M_l}(\vec{i})$
the above assumption does not hold -- e.g. for $l=1$ we need to be able to calculate a Green's function for the hole created at e.g. site $\vec{i}+ (1,0)$ and then annihilated at e.g. site $ \vec{i} - (0,1)$, see Fig.~\ref{fig:rot_cartoon}. 
Hence,
we extend the SAW method to calculate also the off-diagonal coefficients. This is a lengthy exercise that is discussed in detail in Appendix~\ref{sec:appendix:greens_function}-\ref{sec:appendix:diagrams}.  Note that we derive the exact formula for the off-diagonal terms by expressing them through the diagonal terms which we can calculate within SAW. Although the final expression is relatively complex it is not important to study it to understand the results of the main text.
Lastly, the important point to note is that 
in our approach
there are no further approximations beyond the SAW. 

\section{\label{sec:results_arpes}Results}

In the above section we have shown how to efficiently calculate the off-diagonal coefficients of the Greens function within the self-avoiding paths approximation. From the latter we can easily obtain the rotational spectral function $ A_{M_l}(\omega)$, see Eq.~\ref{eq:ag}.
As already mentioned, the primary goal of the paper is to calculate the rotational spectral function for three distinct cases: without rotations $(l=0)$, with first-generation rotational excitation $(l=1)$, and with a second-generation rotational excitation $(l=2)$.
Moreover, in the discussion below we present how the obtained results depend on the choice of the lattice (either the square or Bethe lattice) and on the contribution from the magnon-magnon interactions (either included or discarded). 
Note that all spectral functions are shifted in $\omega$ such that the ground state energy equals 0. 
(Depending on the case there may be a peak at 0 energy or not---for in general the weight at the ground state energy does not have to be finite.)

Note that for the 2D~$t$--$J^z$ model to be at least partially realistic one assumes that it is rather defined on a square lattice than on a corresponding Bethe lattice (with coordination number $z=4$). The $t$--$J^z$~model on a Bethe lattice is a simpler model than the $t$--$J^z$ on a square and therefore should be considered as a simplified and approximate version of the $t$--$J^z$ model on a square lattice. Besides, when expressed in the slave-fermion basis (a.k.a. magnon-holon basis) the $t$--$J^z$~model  always contains the magnon-magnon interactions (of a specific value). Thus, skipping the magnon-magnon interaction leads to an approximation. Altogether, this means that the $t$--$J^z$~model on a square lattice (with the magnon-magnon interactions implicitly included) should be considered as the `main', i.e. most-realistic, model. On the other hand, the other three versions of this model considered here, i.e. the  $t$--$J^z$~model on a Bethe lattice with {\it or} without magnon-magnon interactions, and the $t$--$J^z$ model on the square lattice without magnon-magnon interactions are merely approximate versions of  the $t$--$J^z$~model on a square lattice.

\subsection{\label{sec:results:no_rot}Spectral functions without rotations}

\begin{figure}[t!]
\begin{center}
    \begin{minipage}[c]{\columnwidth}
        standard spectral function\\
        w/o magnon-magnon interactions\\
    	\includegraphics[width=0.49\columnwidth]
    	{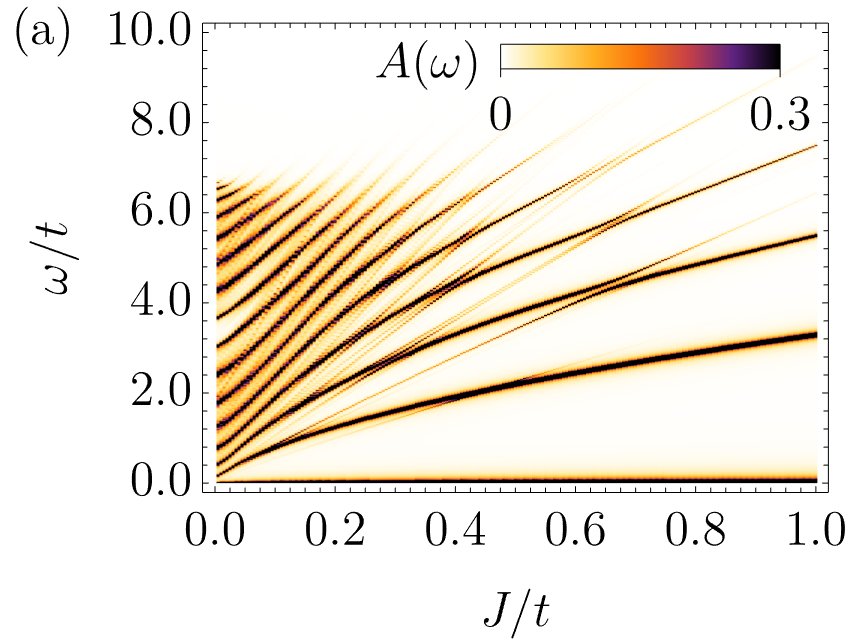}
    	\includegraphics[width=0.49\columnwidth]
    	{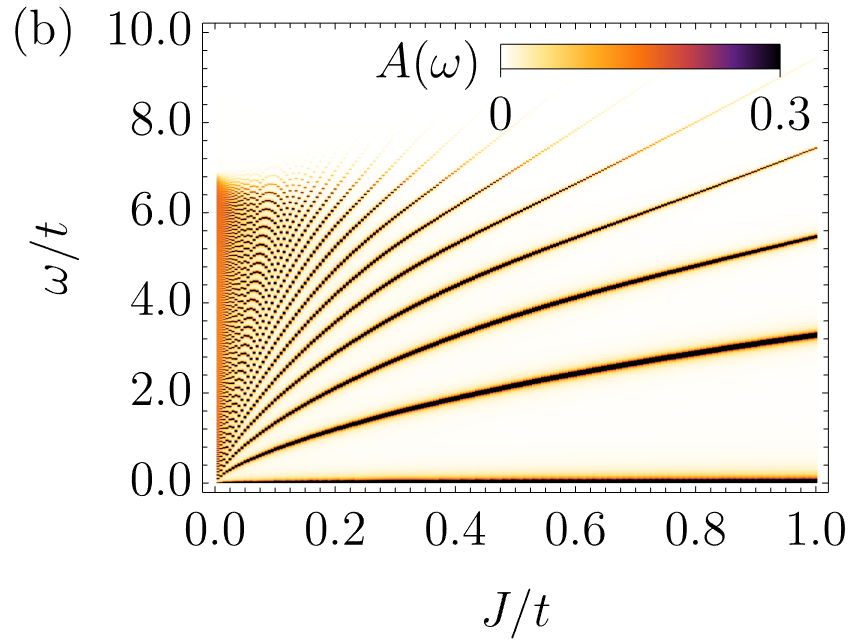}
    \end{minipage}
\end{center}
	\caption{Dependence of the spectral function $A(\omega)$ (zeroth generation) of a single hole in the Ising antiferromagnet on the coupling constant $J/t$. Results obtained {\it without} the magnon-magnon interactions and: (a) on the square lattice and (b) on the Bethe lattice.
	}\label{fig:no_rot_no_mag}
\end{figure}

\begin{figure}[t!]
\begin{center}
    \begin{minipage}[c]{\columnwidth}
        standard spectral function\\
        w/ magnon-magnon interactions\\
        \includegraphics[width=0.49\columnwidth]
	   {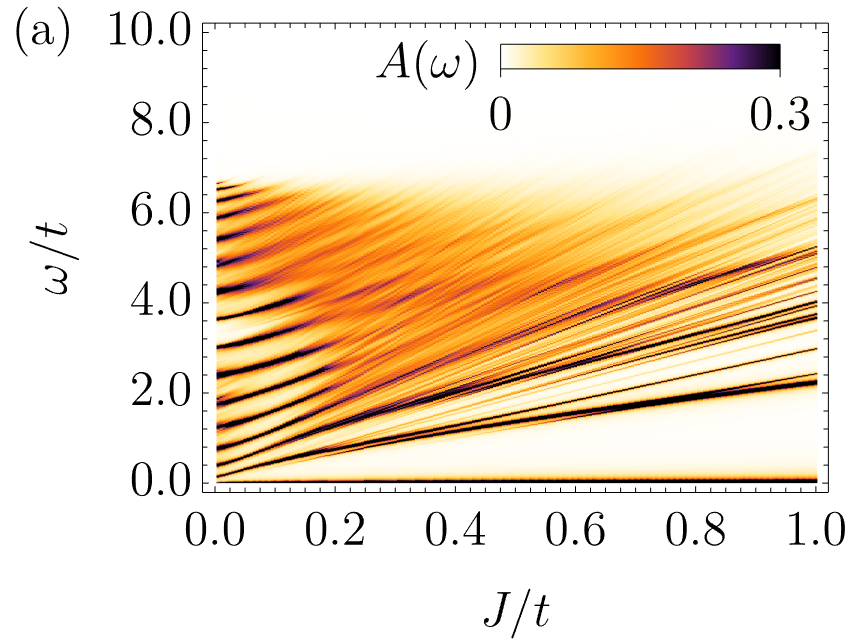}
	   \includegraphics[width=0.49\columnwidth]
	   {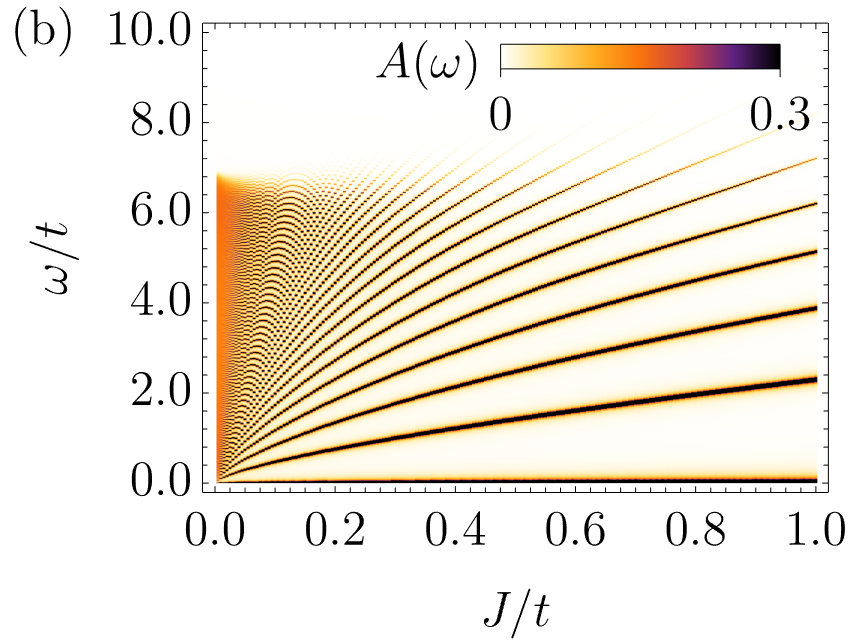}
    \end{minipage}
	\caption{Dependence of the spectral function $A(\omega)$ (zeroth generation) of a single hole in the Ising antiferromagnet on the coupling constant $J/t$. Results obtained after including the magnon-magnon interactions and: (a) on the square lattice and (b) on the Bethe lattice.
	}\label{fig:no_rot}
	\end{center}
\end{figure}

We begin by studying the evolution of the zeroth-generation spectral function {\it without rotations} $A(\omega) = A_{M_0}(\omega)$ -- i.e. the `standard' spectral function, see Fig.~\ref{fig:rot_cartoon}(a) and Eq.~\eqref{eq:R_0}. We plot the respective spectra upon increasing the spin exchange $J/t$ in Fig.~\ref{fig:no_rot_no_mag} and Fig.~\ref{fig:no_rot}. In the simplest case, i.e. without the magnon-magnon interactions included shown in Fig.~\ref{fig:no_rot_no_mag},
we can see that overall the peaks present in the Bethe lattice are well visible also on the square lattice. Distance between these peaks scales as $(J/t)^{2/3}$, as predicted by mapping the lattice problem of a hole in the string potential onto a continuous linear potential. Crucially, however, we also observe a rather subtle, though qualitative, effect when changing the lattice from the Bethe [panel (b)] to the square [panel (a)]: in the latter case we observe additional spectral features that represent eigenstates whose energy position scales {\it linearly} with $J/t$. Since these spectral features are visible also for large $J/t$, they should not be destroyed by including the hole motion in the closed paths, i.e. when going beyond the SAW approximation. In fact, we have unambiguously verified using exact diagonalisation that these features are {\it not} an artifact of the SAW approximation, see Appendix~\ref{sec:appendix:ED} for details.

Thus, we observe that even the simplest case of the hole coupled to the noninteracting magnons on the square lattice shows a qualitatively richer physics than the corresponding Bethe lattice, and therefore goes beyond the `hole in the string potential' and the eigenergies scaling as $(J/t)^{2/3}$ paradigm. 
In fact, the unambiguous detection of eigenstates whose energies scale linearly with $J/t$ is the main result of this paper -- while their apparent presence
in the rotational spectra is further discussed below,
their detailed origin is explained in Sec.~\ref{sec:discussion}.

Before we continue let us leave a side remark related to the apparent discrepancy between the density of peaks for small $J/t < 0.2$ on the square and the Bethe lattices. It is not a genuine result, for it stems from the limitations of the method for small $J/t$. Whereas for the Bethe lattice we are able to include chains of up to 1000 magnons in length, for the square lattice this number is reduced to several magnons (usually slightly less than 20) which is not enough to obtain converged results for $J/t < 0.2$. 
In the end, the number of peaks we observe for small $J/t <0.2$ on the square lattice is too small and in reality the obtained results should be much more similar to the Bethe lattice case.

Next, let us analyze the case with the magnon-magnon interactions included -
this corresponds to the genuine \tjzm{}. Comparing Fig.~\ref{fig:no_rot} (with magnon-magnon interactions) against Fig.~\ref{fig:no_rot_no_mag} (without magnon-magnon interactions) we observe that for the Bethe lattice the magnon-magnon interactions only quantitatively influence the spectral function. In fact, they merely rescale the distance between the peaks. This result is not surprising and it can be easily explained, as discussed in the previous work~\cite{Wrzosek2021}. On the other hand, on the square lattice the magnon-magnon interactions introduce a broad continuum. This result is more subtle and comes from the distinct energies of paths with the same number of magnons. The two paths with the same number of magnons may have different energy if they become tangential, allowing for additional interactions between the magnons, again see~\cite{Wrzosek2021} for details. In an effective string picture of the doped $t$--$J^z$~model~\cite{Gru18} this corresponds to self-interaction of the strings. On the Bethe lattice this is not possible and the magnons always interact  with each other solely along the path of the moving hole~\cite{Bie19, Wrzosek2021}. At the same time one can still see significantly pronounced weight on top of the continuum at the positions corresponding to peaks visible on the Bethe lattice. 

Besides, the spectrum with the magnon-magnon interactions included on the square lattice shows `additional' peaks that are linear in $J/t$ and are similar to the ones discussed above in the case without the magnon interactions on the square geometry [cf. Fig.~\ref{fig:no_rot_no_mag}(a) and Fig.~\ref{fig:no_rot}(a)]. Interestingly the number of these lines is much higher once the magnon-magnon interactions are included. This is consistent with the understanding of the role of the 
magnon-magnon interactions that drives massive splitting of peaks due to the different energy cost of paths with the same number of magnons.

\begin{figure}[t!]
\begin{center}
    \begin{minipage}[c]{\columnwidth}
        single-rotation spectral function, $m^{(1)}=0$\\
        w/o magnon-magnon interactions\\
        \includegraphics[width=0.49\columnwidth]
    	{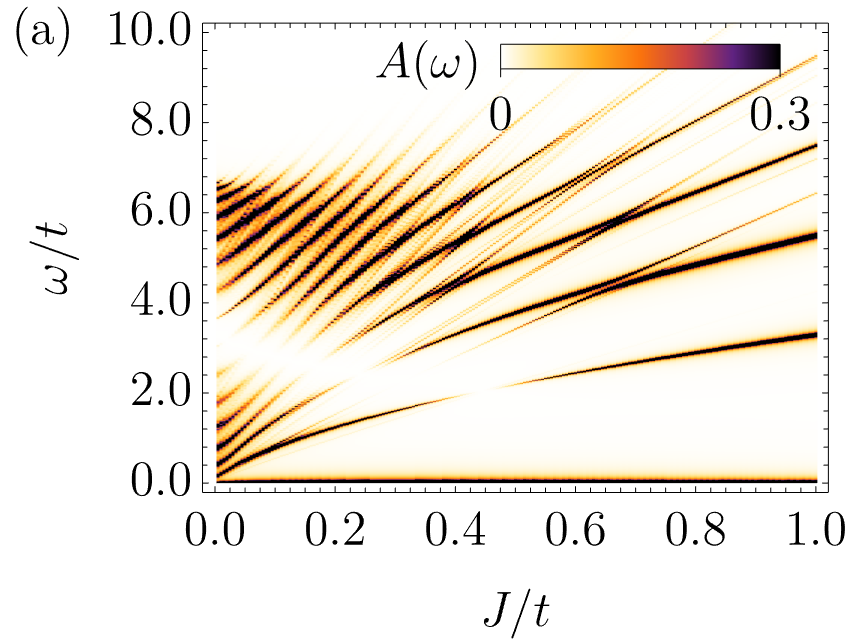}
    	\includegraphics[width=0.49\columnwidth]
    	{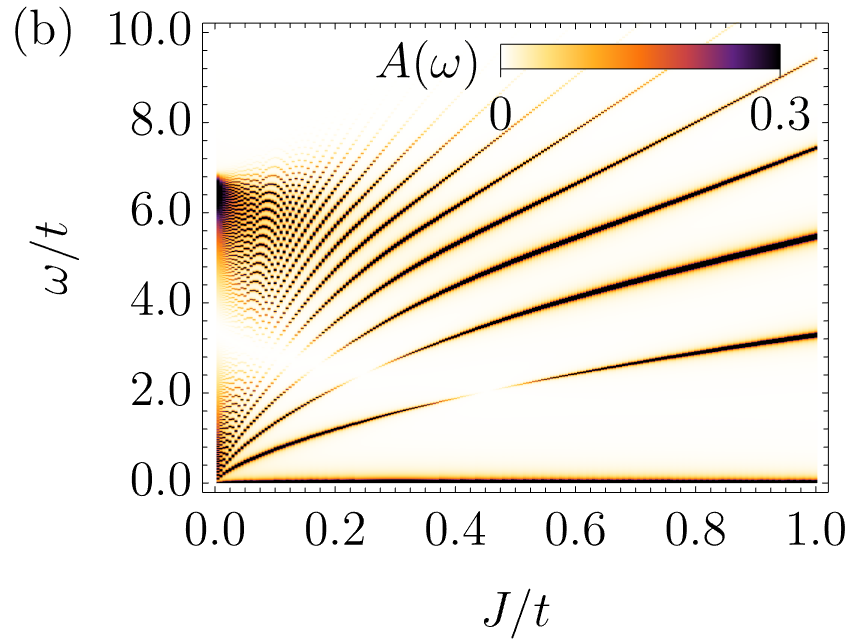}
    \end{minipage}	
\caption{Dependence of the first-generation rotational spectral function $A(\omega)$ with $m^{(1)}=0$ of a single hole in the Ising antiferromagnet on the coupling constant $J/t$. Results obtained {\it without} the magnon-magnon interactions and: (a) on the square lattice and (b) on the Bethe lattice.
}\label{fig:rot_0_no_mag}
\end{center}
\end{figure}

\begin{figure}[t!]
\begin{center}
\begin{minipage}[c]{\columnwidth}
        single-rotation spectral function, $m^{(1)}=0$\\
        w/ magnon-magnon interactions\\
        \includegraphics[width=0.49\columnwidth]
    	{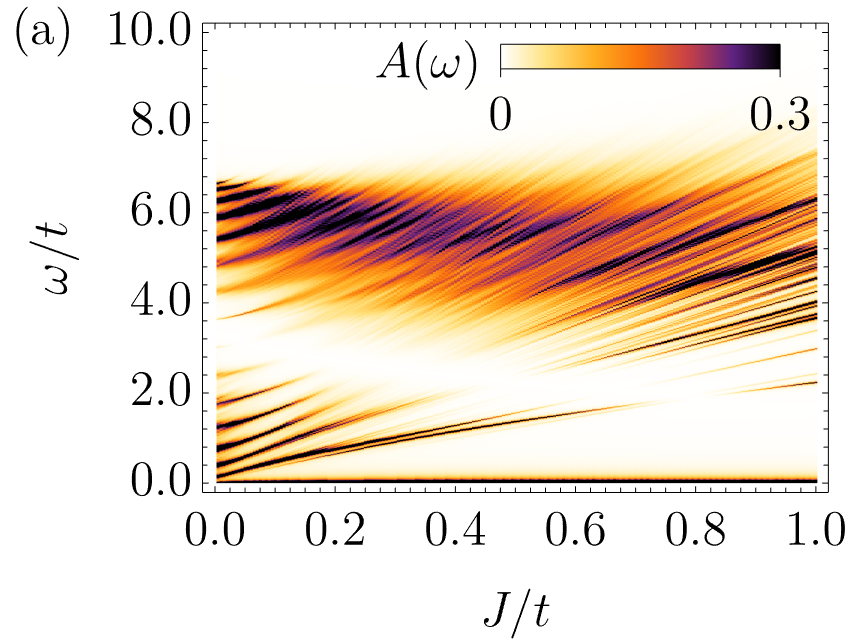}
    	\includegraphics[width=0.49\columnwidth]
    	{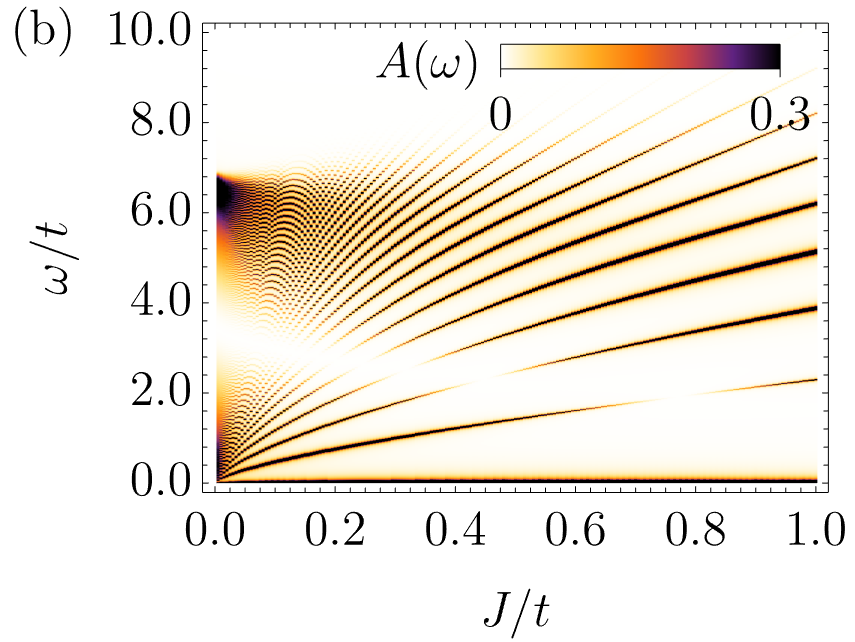}
    \end{minipage}
	\caption{Dependence of the first-generation rotational spectral function $A(\omega)$ with $m^{(1)}=0$ of a single hole in the Ising antiferromagnet on the coupling constant $J/t$. Results obtained after including the magnon-magnon interactions and: (a) on the square lattice and (b) on the Bethe lattice.}\label{fig:rot_0}
\end{center}
\end{figure}

\subsection{\label{sec:results:one_rot} First-generation rotational spectral functions}

Now, let us investigate the spectrum of a single hole with one rotational degree of freedom -- the first-generation rotational spectral function. Microscopically it corresponds to creating a hole in a given site and then propagating it to the nearest sites (i.e. $l=1$) with a phase factor dependent on the direction, see Fig.~\ref{fig:rot_cartoon}(b) and Eq.~\eqref{eq:R_1}. The only free parameter is a phase factor $ m^{(1)} \in \{0,...,3\}$. Note that within the SAW approximation there are only two distinct spectral functions -- one for $m^{(1)} = 0$ and the other one for the remaining values $m^{(1)} = 1,2,3$ which are identical.

We start with the $m^{(1)} = 0$ case. When we compare the spectral functions in Fig.~\ref{fig:rot_0_no_mag} and Fig.~\ref{fig:rot_0} with their counterparts without rotations, Fig.~\ref{fig:no_rot_no_mag} and Fig.~\ref{fig:no_rot} respectively, we observe the same set of peaks. The only qualitative difference between these spectral functions is the missing weight in the middle of the spectrum for all the cases with rotational degree of freedom. Rotational degree of freedom is necessary but not sufficient condition for this effect to appear. The physical origin of the strongly suppressed spectral weight, observed in all excited states but not in the ground state, is the presence of a node in the radial part of the wavefunction. We will discuss this effect later in more details. Due to the summation rules, the missing weight in the middle of the spectrum, makes the peaks more pronounced everywhere else. Altogether, the first-generation rotational spectral functions with $m^{(1)} = 0$ do not seem to provide any additional information over the zeroth-generation spectral functions without rotations.

\begin{figure}[t!]
\begin{center}
    \begin{minipage}[c]{\columnwidth}
        single-rotation spectral function, $m^{(1)}=1$\\
        w/o magnon-magnon interactions\\
    	\includegraphics[width=0.49\columnwidth]
    	{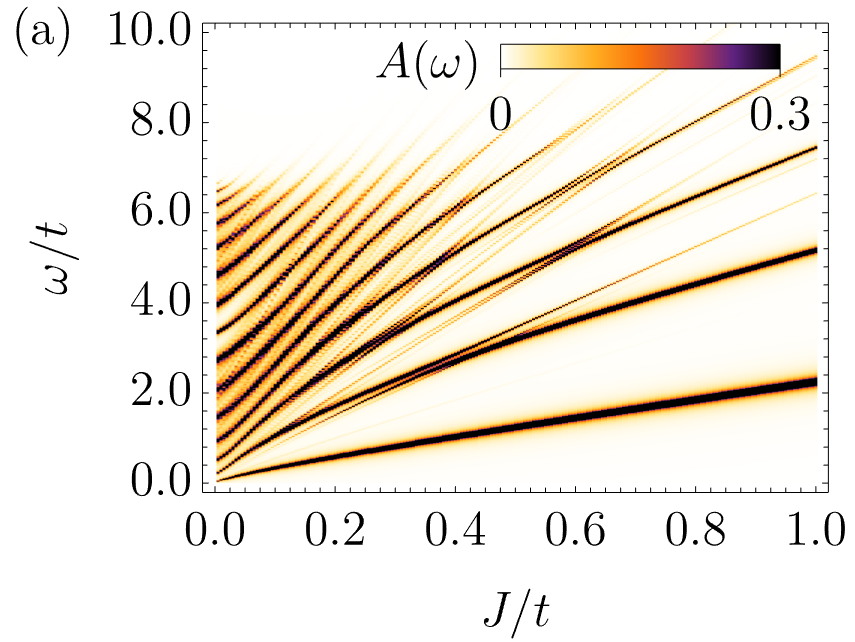}
    	\includegraphics[width=0.49\columnwidth]
    	{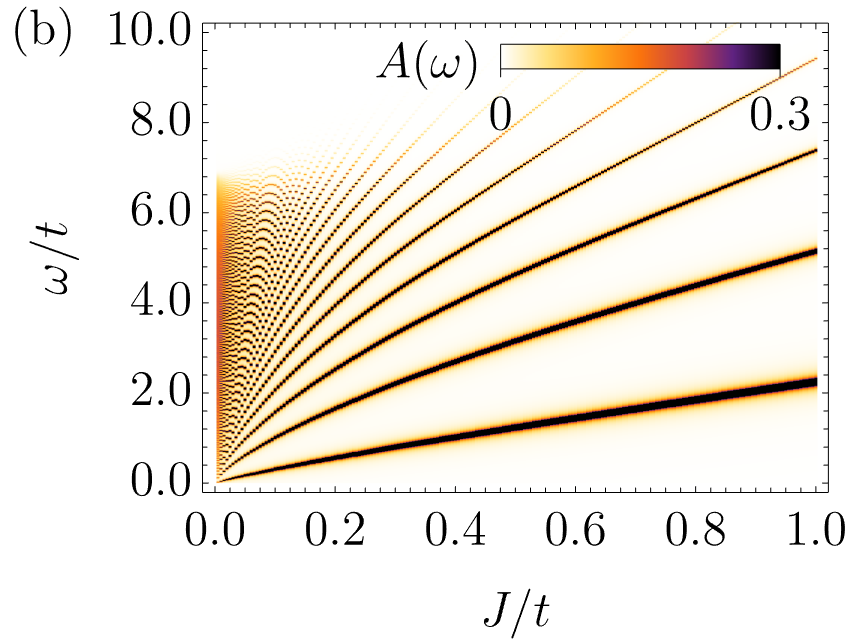}
    \end{minipage}
	\caption{Dependence of the first-generation rotational spectral function $A(\omega)$ with $m^{(1)}=1$ of a single hole in the Ising antiferromagnet on the coupling constant $J/t$. Results obtained {\it without} the magnon-magnon interactions and: (a) on the square lattice and (b) on the Bethe lattice.
}\label{fig:rot_1_no_mag}
\end{center}
\end{figure}

\begin{figure}[t!]
\begin{center}
    \begin{minipage}[c]{\columnwidth}
        single-rotation spectral function, $m^{(1)}=1$\\
        w/ magnon-magnon interactions\\
    	\includegraphics[width=0.49\columnwidth]
    	{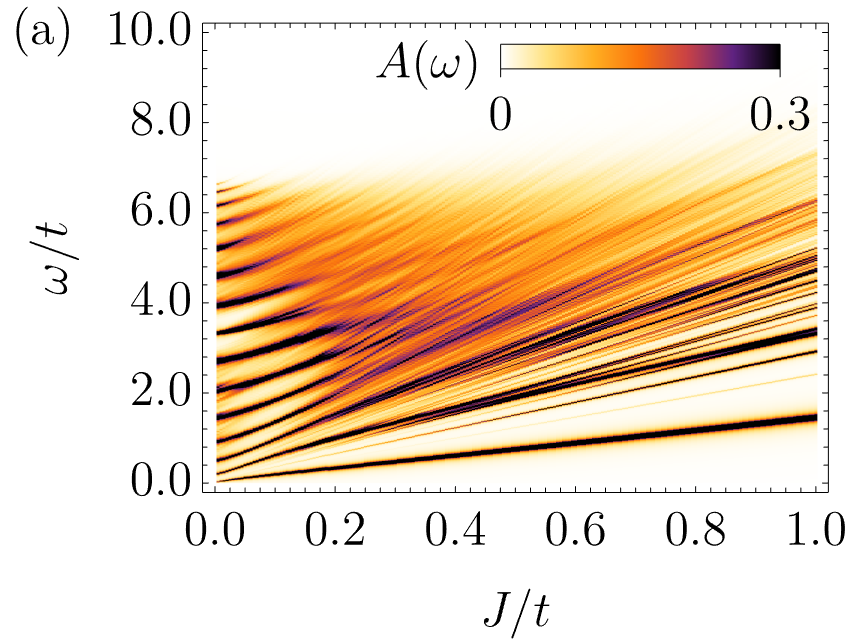}
    	\includegraphics[width=0.49\columnwidth]
    	{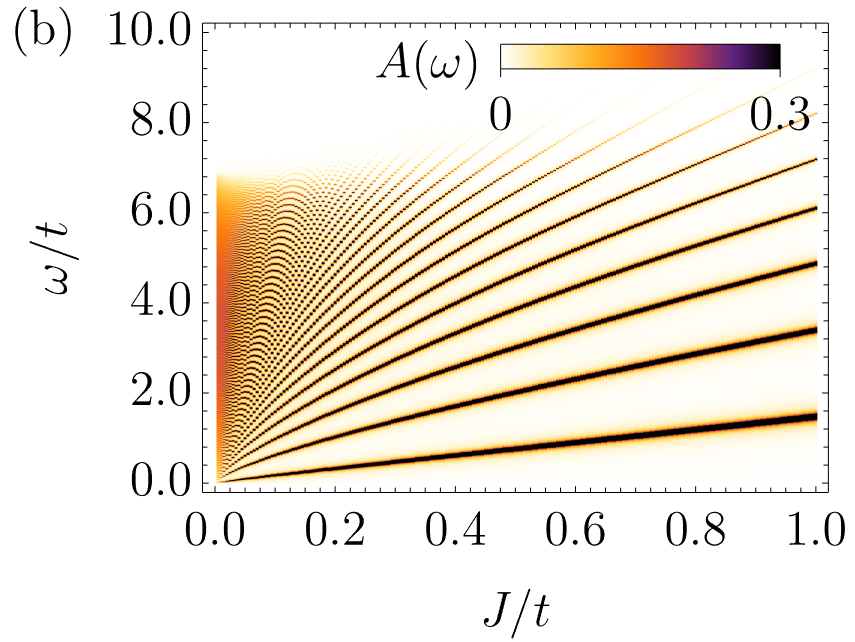}
    \end{minipage}
	\caption{Dependence of the first-generation rotational spectral function $A(\omega)$ with $m^{(1)}=1$ of a single hole in the Ising antiferromagnet on the coupling constant $J/t$. Results obtained after including the magnon-magnon interactions and: (a) on the square lattice and (b) on the Bethe lattice.}\label{fig:rot_1}
	\end{center}
\end{figure}

Next we analyse the case with $m^{(1)} = 1,2,3$ in Figs. ~\ref{fig:rot_1_no_mag}, \ref{fig:rot_1}. 
While in general 
we observe similar behavior as in the previously discussed spectra,
such as the $(J/t)^{2/3}$ and $J/t$ scalings on the sqaure lattice or onset of continua due to the magnon-magnon interactions,
there are also subtle differences:

We can see that the ground state is not visible in these cases -- there is no peak at zero energy neither in Fig.~\ref{fig:rot_1_no_mag} nor in Fig.~\ref{fig:rot_1}. Moreover, for the Bethe lattice case we see that {\it none} of the peaks visible in the previous spectra is present in the spectra with $m^{(1)} = 1$. For the square lattice a similar effect can also be observed.

There is one more difference between the  $m^{(1)} = 1,2,3$ first-generation rotational spectrum and all of the previous spectra (the $m^{(1)}=0$ first-generation rotational case or the `standard', zeroth-generation spectrum).
Namely, not only the peaks are at different positions compared to the previous spectra but also they behave differently when $J/t$ varies. We can clearly observe that the lowest peak in Fig.~\ref{fig:rot_1_no_mag} and Fig.~\ref{fig:rot_1}, both with and without the magnon-magnon interactions, is linear in $J/t$. Higher energy peaks seem to have additional non-linear component but they are also close to the linear dependence, especially for large $J/t$. On the other hand, there is no missing weight in the middle of the spectra, despite having rotational degree of freedom. Thus, one can see that rotations are not sufficient to create the gap in the spectrum by redistributing the weight to the lower and higher energies.

\subsection{\label{sec:results:two_rot} Second-generation rotational spectral functions}

The last set of results concerns the spectral functions of a hole with two rotational degrees of freedom -- the second-generation rotational spectral function. It corresponds to creating a hole in a given site and then propagating it twice (without returning), i.e. to next nearest neighbour sites (i.e. $l=2$) with a phase factor dependent on the direction, see Fig.~\ref{fig:rot_cartoon}(c) and Eq.~\eqref{eq:R_2}. This time there are two free parameters, $ m^{(1)} \in \{0,...,3\}$ and $ m^{(2)} \in \{0,1,2\}$.

\begin{figure}[t!]
\begin{center}
    \begin{minipage}[c]{\columnwidth}
        double-rotation spectral function, $m^{(1)}=m^{(2)}=0$\\
        w/o magnon-magnon interactions\\
    	\includegraphics[width=0.49\columnwidth]
    	{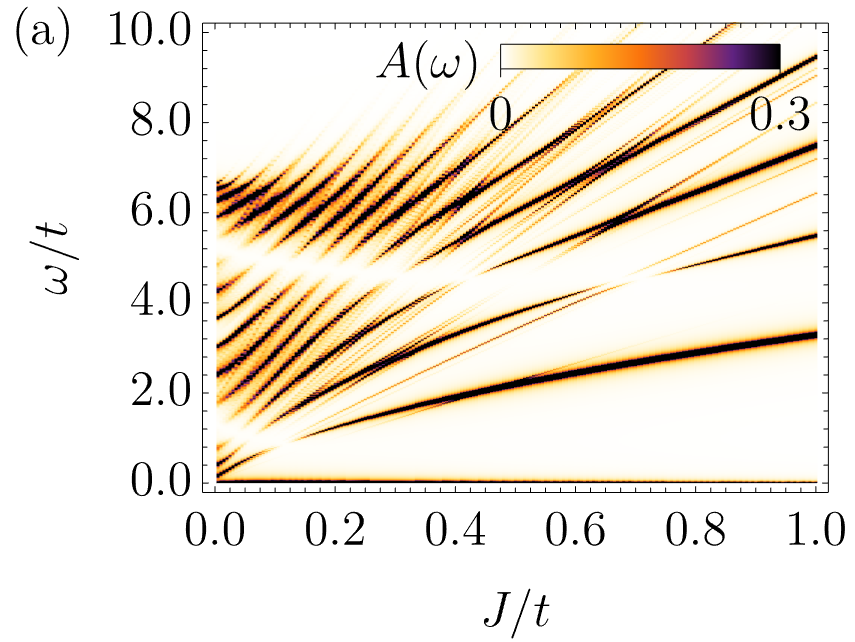}
    	\includegraphics[width=0.49\columnwidth]
    	{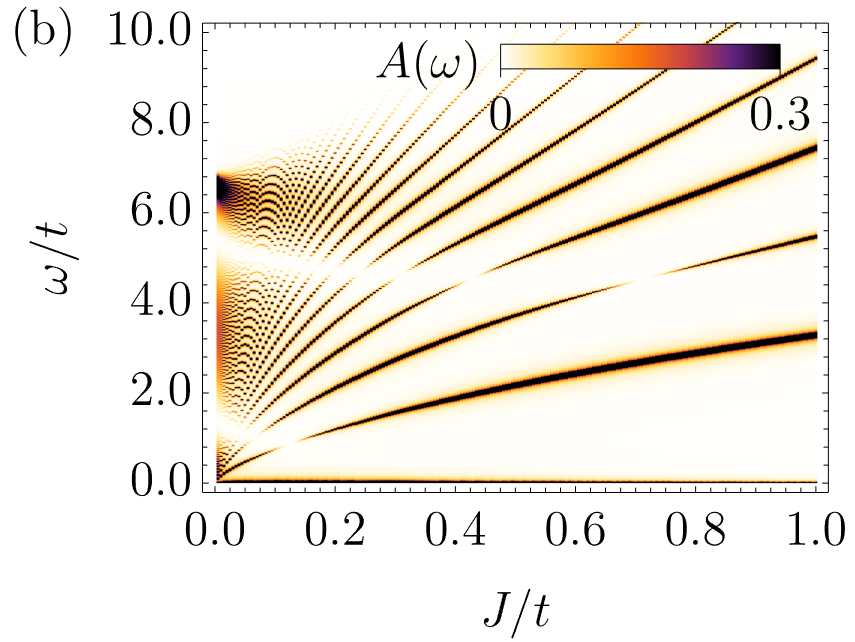}
    \end{minipage}
	\caption{Dependence of the second-generation rotational spectral function $A(\omega)$ with ($m^{(1)}=0, m^{(2)}=0$) of a single hole in the Ising antiferromagnet on the coupling constant $J/t$. Results obtained {\it without} the magnon-magnon interactions and: (a) on the square lattice and (b) on the Bethe lattice.
	}\label{fig:rot_00_no_mag}
\end{center}
\end{figure}

\begin{figure}[t!]
\begin{center}
    \begin{minipage}[c]{\columnwidth}
        double-rotation spectral function, $m^{(1)}=m^{(2)}=0$\\
        w/ magnon-magnon interactions\\
    	\includegraphics[width=0.49\columnwidth]
    	{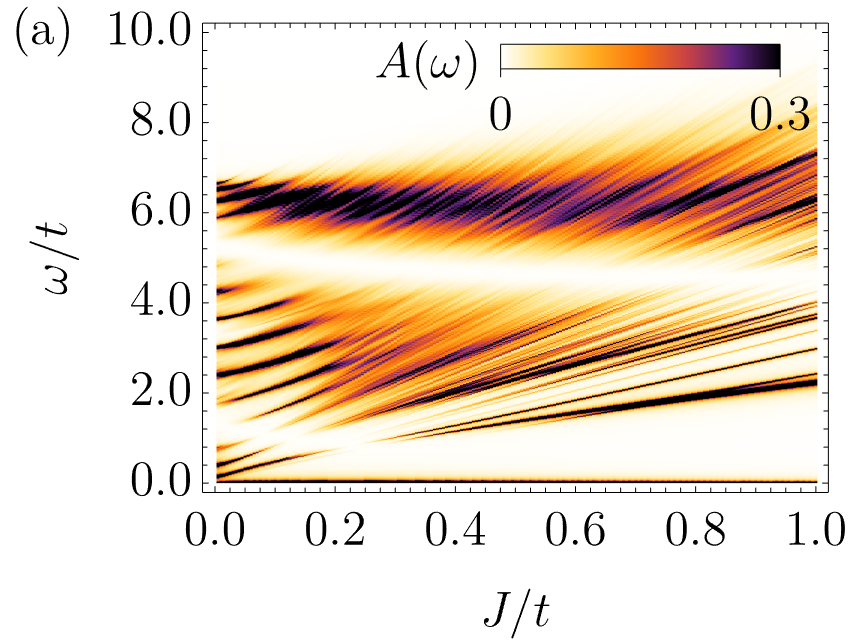}
    	\includegraphics[width=0.49\columnwidth]
    	{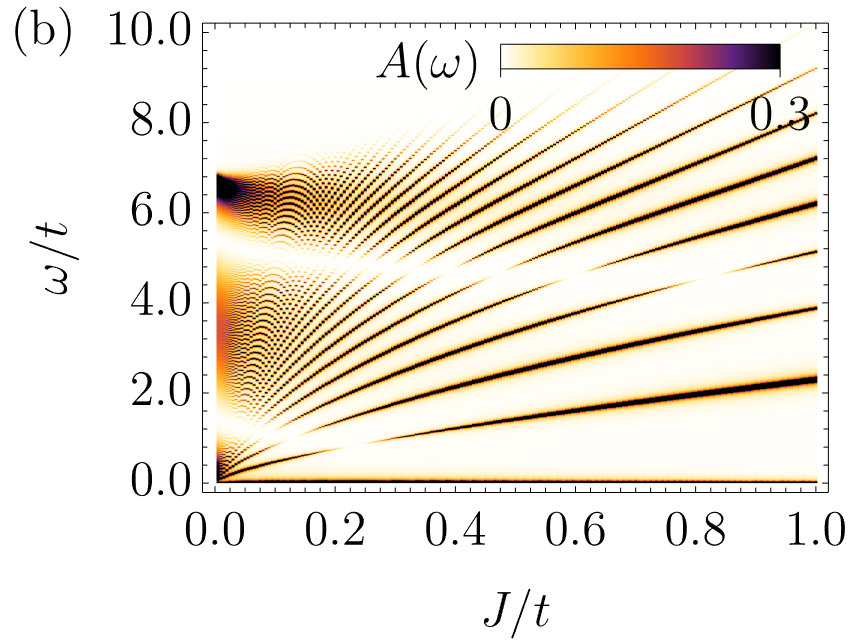}
    \end{minipage}
	\caption{Dependence of the second-generation rotational spectral function $A(\omega)$ with ($m^{(1)}=0, m^{(2)}=0$) of a single hole in the Ising antiferromagnet on the coupling constant $J/t$. Results obtained after including the magnon-magnon interactions and: (a) on the square lattice and (b) on the Bethe lattice.
	}\label{fig:rot_00}
\end{center}
\end{figure}

We start our analysis with the case $(m^{(1)},m^{(2)}) = (0,0)$, i.e. when the hole does not acquire any angular momentum in any of the moves. We can clearly see that the corresponding spectral functions without the magnon-magnon interactions presented in Fig.~\ref{fig:rot_00_no_mag} are qualitatively similar to those of Fig.~\ref{fig:rot_0_no_mag} (and therefore, except for the onset of the washed-away features, also to Fig.~\ref{fig:no_rot_no_mag}). The only difference is that now there are two regions where the weight of the peaks is strongly suppressed. But all the peaks are in exactly same positions in all mentioned figures. The same applies to the cases with the magnon-magnon interactions included ({\it cf}. Fig.~\ref{fig:rot_00}, Fig.~\ref{fig:no_rot} and Fig.~\ref{fig:rot_0}).

\begin{figure}[t!]
\begin{center}
    \begin{minipage}[c]{\columnwidth}
        double-rotation spectral function, $m^{(1)}=1, m^{(2)}=0$\\
        w/o magnon-magnon interactions\\
    	\includegraphics[width=0.49\columnwidth]
    	{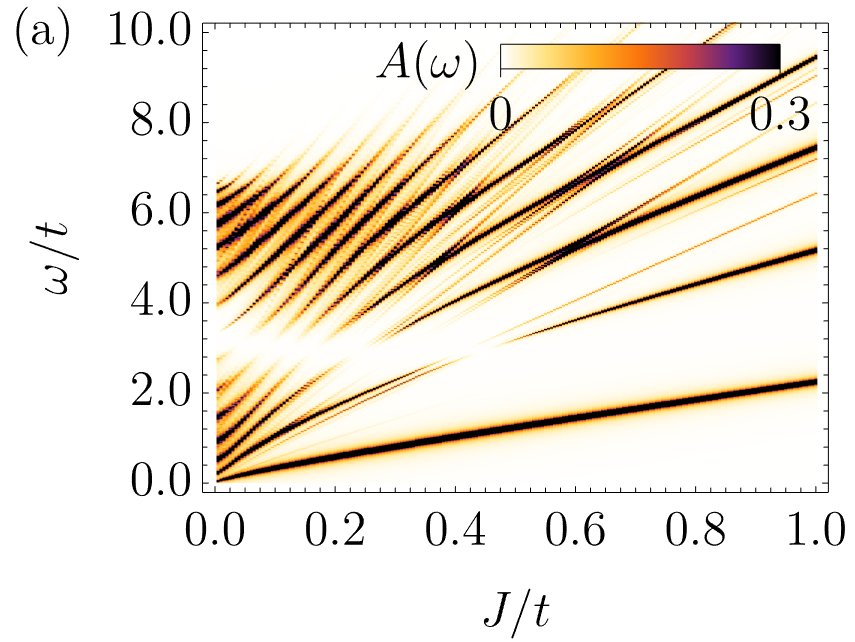}
    	\includegraphics[width=0.49\columnwidth]
    	{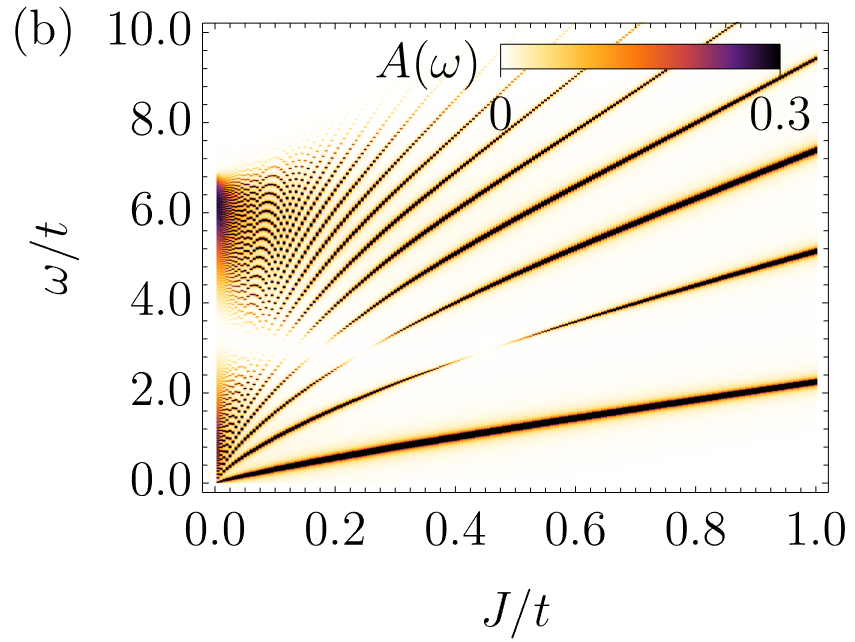}
    \end{minipage}
	\caption{Dependence of the second-generation rotational spectral function $A(\omega)$ with ($m^{(1)}=1, m^{(2)}=0$) of a single hole in the Ising antiferromagnet on the coupling constant $J/t$. Results obtained {\it without} the magnon-magnon interactions and: (a) on the square lattice and (b) on the Bethe lattice.
	}\label{fig:rot_10_no_mag}
\end{center}
\end{figure}

\begin{figure}[t!]
\begin{center}
    \begin{minipage}[c]{\columnwidth}
        double-rotation spectral function, $m^{(1)}=1, m^{(2)}=0$\\
        w/ magnon-magnon interactions\\
    	\includegraphics[width=0.49\columnwidth]
    	{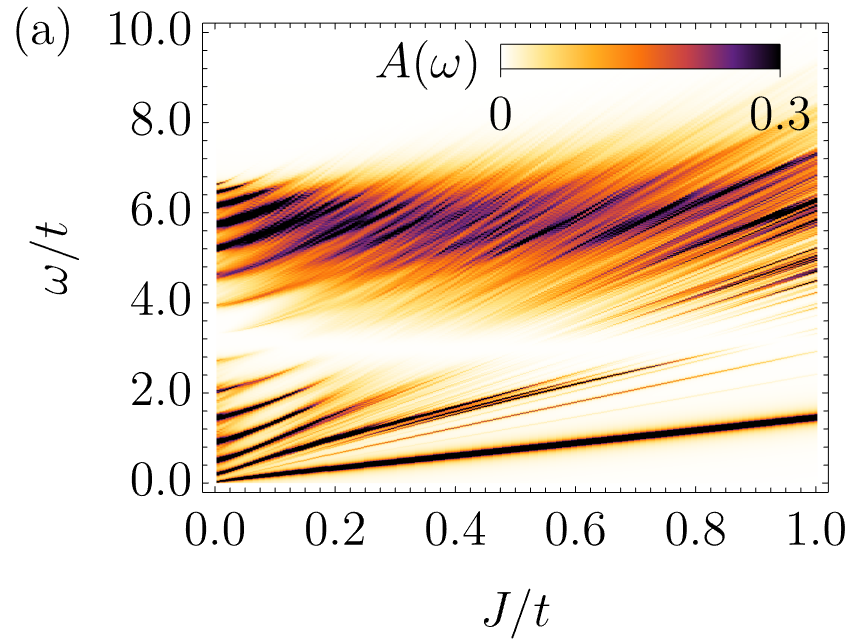}
    	\includegraphics[width=0.49\columnwidth]
    	{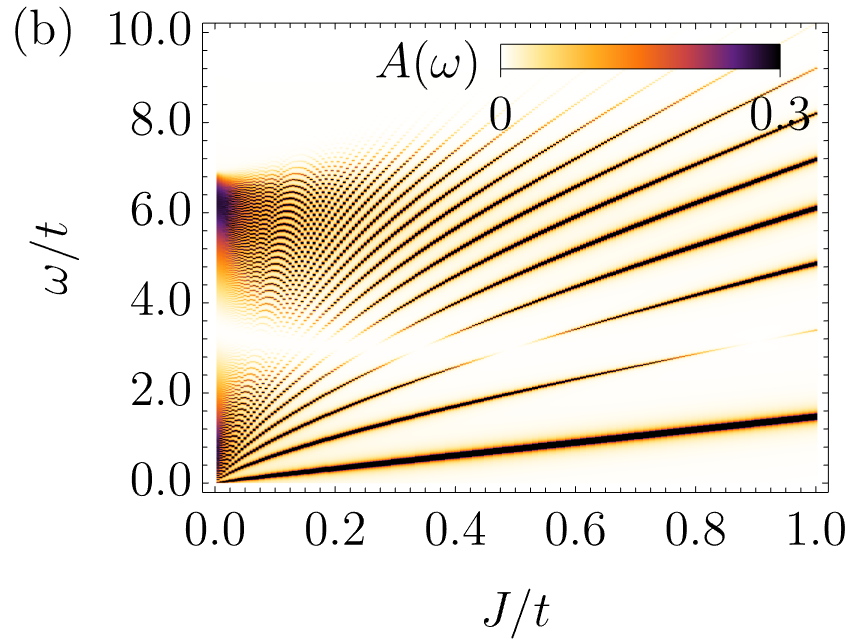}
    \end{minipage}
	\caption{Dependence of the second-generation rotational spectral function $A(\omega)$ with ($m^{(1)}=1, m^{(2)}=0$) of a single hole in the Ising antiferromagnet on the coupling constant $J/t$. Results obtained after including the magnon-magnon interactions and: (a) on the square lattice and (b) on the Bethe lattice.
	}\label{fig:rot_10}
\end{center}
\end{figure}

Next we study the case with $(m^{(1)}, m^{(2)}) = (1,0)$. This case is more interesting. On the qualitative level, i.e. when it comes to the set of peaks present in the spectral function, the spectral functions presented in Fig.~\ref{fig:rot_10_no_mag} are equivalent to previous case of single rotation $m^{(1)} = 1$, see Fig.~\ref{fig:rot_1_no_mag}. The same applies when we introduce the magnon-magnon interactions, cf. Fig.~\ref{fig:rot_10} with Fig.~\ref{fig:rot_1}. On top of this, the fact that now we have $m^{(2)} = 0$ introduces a region where the weight of the peaks is removed. We could see the same behaviour with $m^{(1)} = 0$ in the first-generation spectral functions (see Fig.~\ref{fig:rot_0_no_mag}, Fig.~\ref{fig:rot_0}).

\begin{figure}[t!]
\begin{center}
    \begin{minipage}[c]{\columnwidth}
        double-rotation spectral function, $m^{(1)}=0, m^{(2)}=1$\\
        w/o magnon-magnon interactions\\
    	\includegraphics[width=0.49\columnwidth]
    	{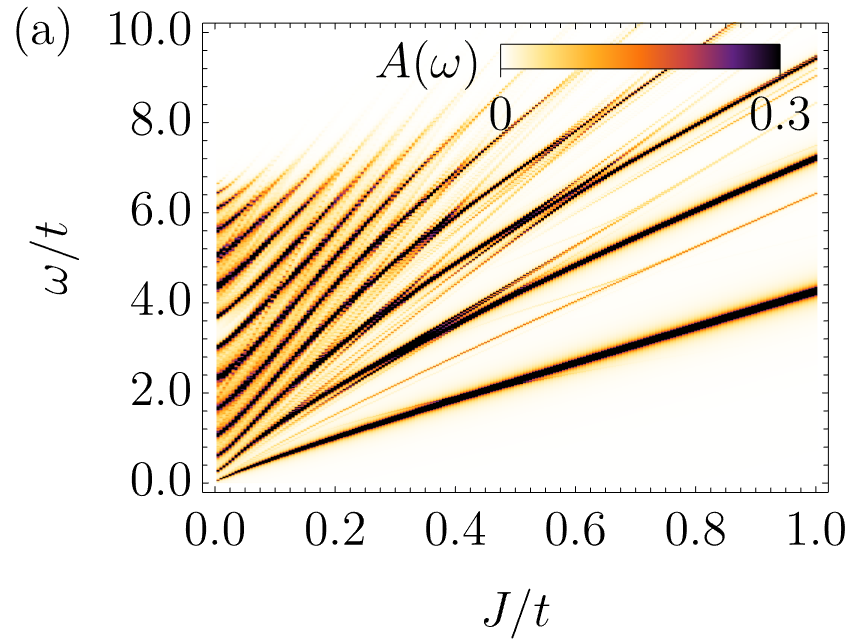}
    	\includegraphics[width=0.49\columnwidth]
    	{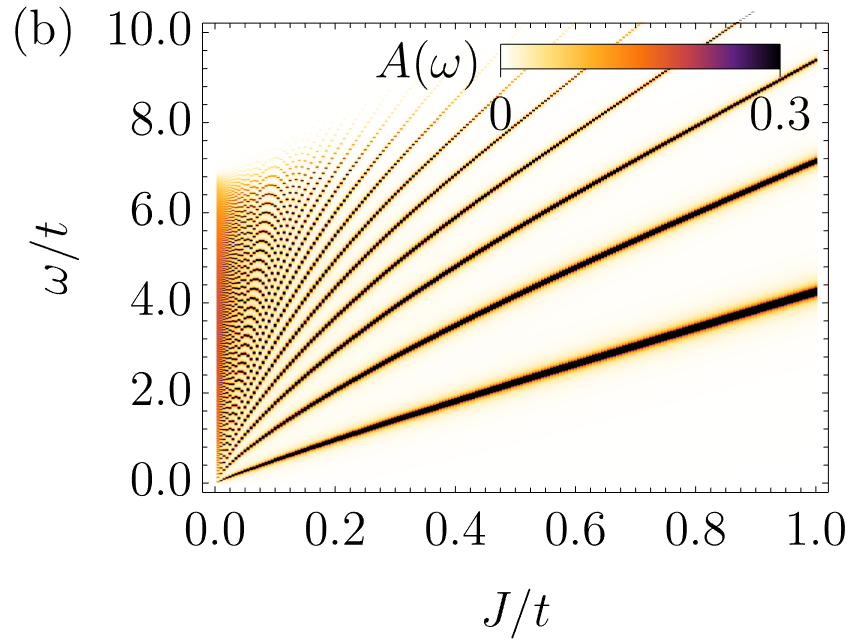}
    \end{minipage}
	\caption{Dependence of the second-generation rotational spectral function $A(\omega)$ with ($m^{(1)}=0, m^{(2)}=1$) of a single hole in the Ising antiferromagnet on the coupling constant $J/t$. Results obtained {\it without} the magnon-magnon interactions and: (a) on the square lattice and (b) on the Bethe lattice.}
	\label{fig:rot_01_no_mag}
	\end{center}
\end{figure}

\begin{figure}[t!]
\begin{center}
    \begin{minipage}[c]{\columnwidth}
        double-rotation spectral function, $m^{(1)}=0, m^{(2)}=1$\\
        w/ magnon-magnon interactions\\
    	\includegraphics[width=0.49\columnwidth]
    	{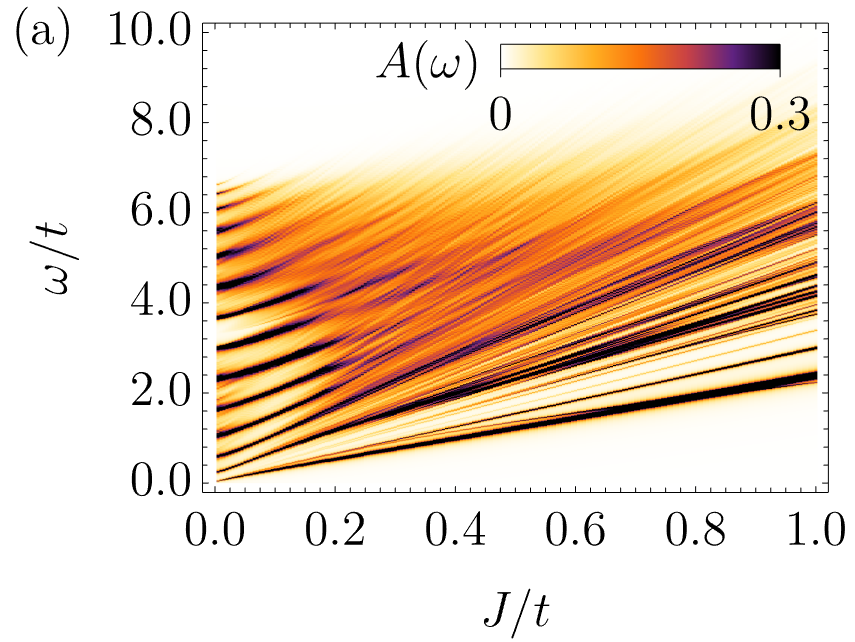}
    	\includegraphics[width=0.49\columnwidth]
    	{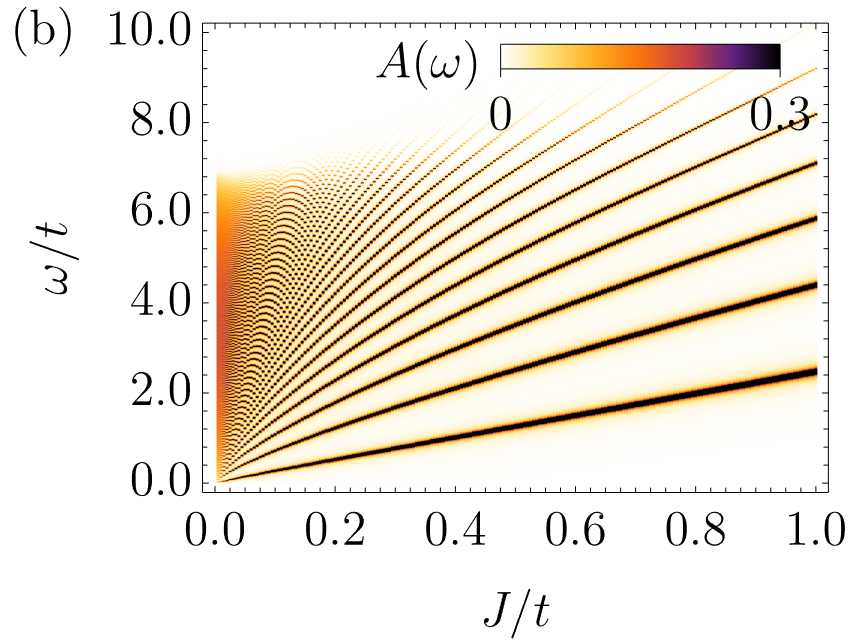}
    \end{minipage}
	\caption{Dependence of the second-generation rotational spectral function $A(\omega)$ with ($m^{(1)}=0, m^{(2)}=1$) of a single hole in the Ising antiferromagnet on the coupling constant $J/t$. Results obtained after including the magnon-magnon interactions and: (a) on the square lattice and (b) on the Bethe lattice.}\label{fig:rot_01}
\end{center}
\end{figure}

Let us now swap the values of the phase factor and analyse the case of $(m^{(1)},m^{(2)}) = (0,1)$. Corresponding second-generation spectral functions omitting (including) the magnon-magnon interactions are presented in Fig.~\ref{fig:rot_01_no_mag} (Fig.~\ref{fig:rot_01}), respectively. We observe that there is no region of suppressed weight, despite that $m^{(1)} = 0$. Combining this with the previous results leads us to the conjecture that the number of regions with damped weight is equal to the number of consecutive jumps of the hole without acquiring the angular momentum counted from the last move to the first move. 

Again, the set of peaks present in Fig.~\ref{fig:rot_01_no_mag} and Fig.~\ref{fig:rot_01} is on the qualitative level similar to the case with the single rotation and $m^{(1)} = 1$, (cf. Fig.~\ref{fig:rot_1_no_mag} and Fig.~\ref{fig:rot_1}). Specifically, there is no peak at the ground state energy and the first visible peak is linear in $J/t$ and several energy peaks are linear in $J/t$. Quantitatively, however, the peak positions in Figs.~\ref{fig:rot_01_no_mag}, \ref{fig:rot_01} [for $m^{(2)}=1$] differ from those in Figs.~\ref{fig:rot_1_no_mag}, \ref{fig:rot_1} [for $m^{(1)}=1$ but no $m^{(2)}$].

\begin{figure}[t!]
\begin{center}
    \begin{minipage}[c]{\columnwidth}
        double-rotation spectral function, $m^{(1)}=m^{(2)}=1$\\
        w/o magnon-magnon interactions\\
    	\includegraphics[width=0.49\columnwidth]
    	{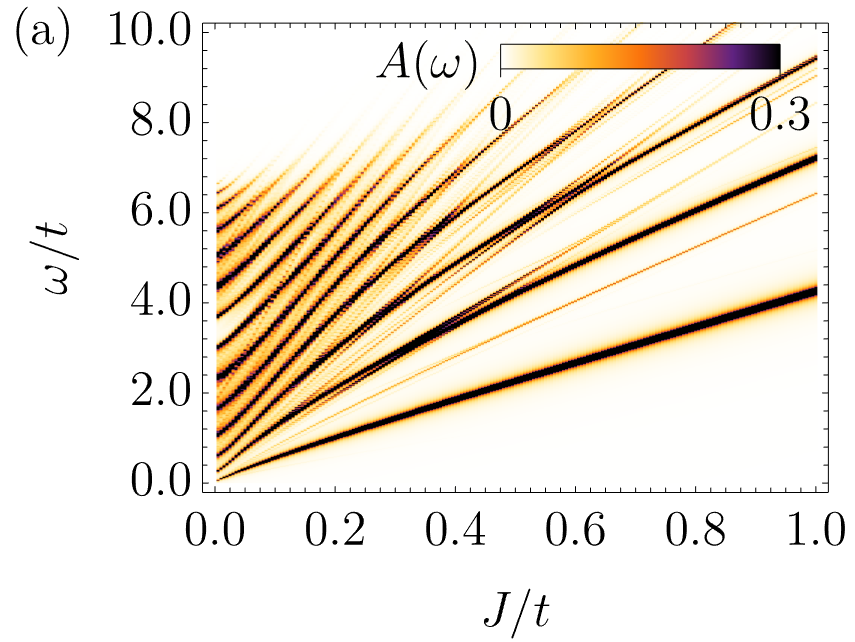}
    	\includegraphics[width=0.49\columnwidth]
    	{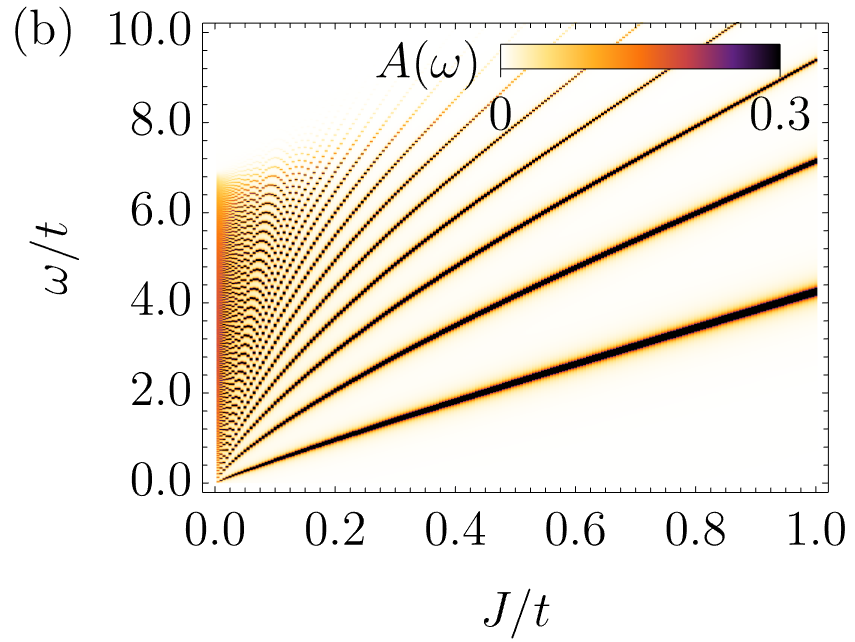}
    \end{minipage}
	\caption{Dependence of the second-generation rotational spectral function $A(\omega)$ with ($m^{(1)}=1, m^{(2)}=1$) of a single hole in the Ising antiferromagnet on the coupling constant $J/t$. Results obtained {\it without} the magnon-magnon interactions and: (a) on the square lattice and (b) on the Bethe lattice.}\label{fig:rot_11_no_mag}
	\end{center}
\end{figure}

\begin{figure}[t!]
\begin{center}
    \begin{minipage}[c]{\columnwidth}
        double-rotation spectral function, $m^{(1)}=m^{(2)}=1$\\
        w/ magnon-magnon interactions\\
    	\includegraphics[width=0.49\columnwidth]
    	{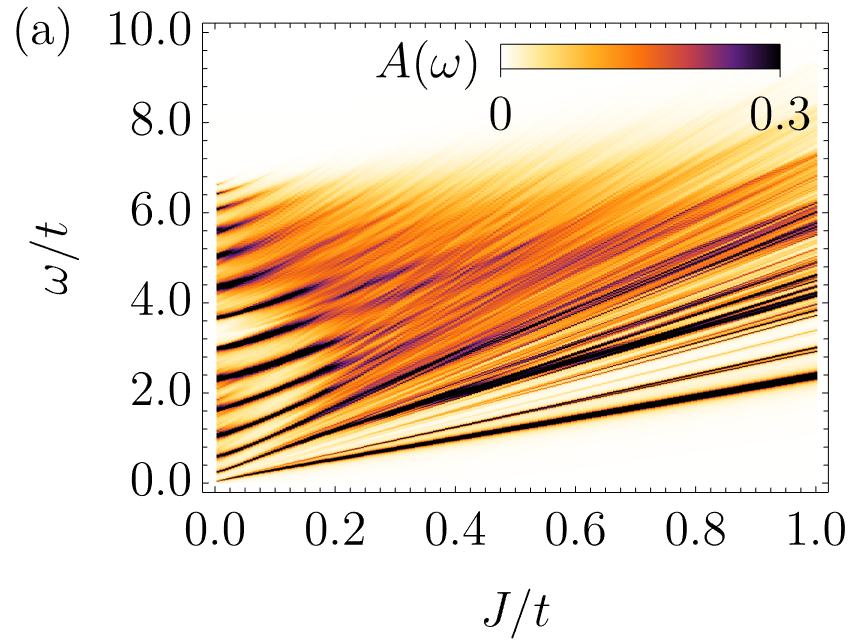}
    	\includegraphics[width=0.49\columnwidth]
    	{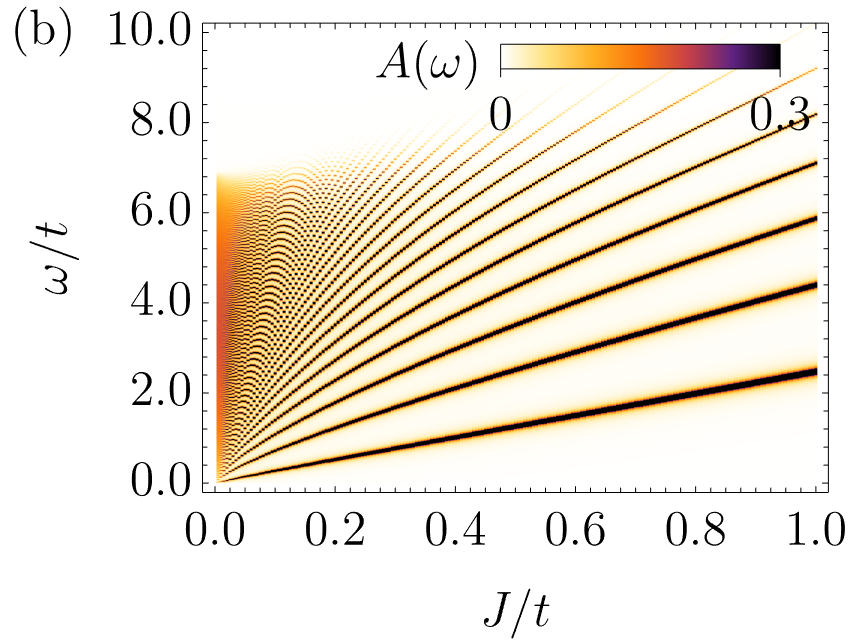}
    \end{minipage}
	\caption{Dependence of the second-generation rotational spectral function $A(\omega)$ with ($m^{(1)}=1, m^{(2)}=1$) of a single hole in the Ising antiferromagnet on the coupling constant $J/t$. Results obtained after including the magnon-magnon interactions and: (a) on the square lattice and (b) on the Bethe lattice.}\label{fig:rot_11}
\end{center}
\end{figure}

The last of the presented set of figures concerns $(m^{(1)},m^{(2)}) = (1,1)$. We can see that result for the Bethe lattice [Fig.~\ref{fig:rot_11_no_mag}(b) and Fig.~\ref{fig:rot_11}(b)] looks the same as in the case of $m^{(1)} = 0$ discussed in the previous paragraphs [see Fig.~\ref{fig:rot_01_no_mag}(b) and Fig.~\ref{fig:rot_01}(b)]. Similarly, results on the square lattice seem to be the same -- but in this case there are small differences, barely visible in the spectrum. For example, compared to the $m^{(1)} = 0$ case, an extremely faint additional peak appears at the lowest energies, even without the magnon-magnon interactions [{\it cf}. Fig.~\ref{fig:rot_11_no_mag}(a) and Fig.~\ref{fig:rot_01_no_mag}(a)]. When the magnon-magnon interactions are included these changes in the fine structure of the spectrum are more pronounced [{\it cf}. Fig.~\ref{fig:rot_11}(a) and Fig.~\ref{fig:rot_01}(a)]. Such behaviour suggests that this effect originates from the existence of tangential paths on the square lattice. Nevertheless, it is a rather subtle contribution to the overall shape of the spectrum.

\section{\label{sec:discussion}
Discussion:\\ origin of  eigenenergies $\propto J/t$}

To simplify the study below, we neglect the magnon-magnon interactions in this section.  Note that, as discussed above, the onset
eigenstates of the problem whose energies scale $\propto J/t$ occurs both for the case with and without magnon-magnon interactions.

\subsection{Preliminaries:
two equivalent bases on a Bethe lattice}
Let us consider the $t$--$J^z$ Hamiltonian $\hat{H}$ (in holon-magnon basis) and an empty (for holes and magnons) Bethe lattice with coordination $z=4$. 
Let us denote this empty lattice as vacuum state $\ket{\varnothing}$. Next, let us choose a certain site in the Bethe lattice and refer to this site as the origin of the lattice by denoting it as $j=0$. Let us further create a hole at the origin of the lattice and let us denote the resulting \textit{initial} state as $\ket{0}$,
\begin{equation}
	\ket{0} = \hat{h}_{j=0}^\dag \ket{\varnothing}.
\end{equation} 
Let us consider two orthogonal states $\psi_1$ and $\psi_2$. We will say that $\psi_2$ is $n$-reachable from $\psi_1$ if $n$ is the smallest positive integer such that $\bra{\psi_2}\hat{H}^n\ket{\psi_1} \neq 0$. If such an integer does not exist, we will say that $\psi_2$ and $\psi_1$ are disconnected (or non-reachable from one another). Let us find all the states $n$-reachable from $\ket{0}$ for $n=1,2,3,\hdots$, by acting on $\ket{0}$ with $\hat{H}^n$ for different values of $n$. 

Let us consider $n=1$ at the beginning,
\begin{equation}
	\hat{H}\ket{0} = E_0 \ket{0} - t(\ket{1;0}+\ket{1;1}+\ket{1;2}+\ket{1;3}).
\end{equation}
In the above notation $\ket{n;d_1,...,d_n}$ is a \textit{position} state with $n$ magnons where the hole has moved through bonds in directions $d_1,...,d_n$ (without returning) in the given order (i.e. from $d_1$ to $d_n$). The condition for states to be $n$-reachable does not specify the basis. We could for example choose 4 position states,
\begin{equation}
	\ket{1;0}, \quad \ket{1;1}, \quad \ket{1;2}, \quad \ket{1;3}
\end{equation}
to describe the subspace of states 1-reachable from $\ket{0}$. But we could also mix them to get the so-called \textit{rotational} states -- i.e. for the above example we would then get:
\begin{equation}
	\sket{1;m^{(1)}}_r = \frac{1}{\sqrt{4}}\sum_{d_1=0}^3 \exp\left(\frac{2 \pi i}{4}d_1 m^{(1)}\right)\ket{1;d_1},
\end{equation}
where $m^{(1)} = 0,...,3$. The initial state in the rotational representation is simply $\ket{0}_r = \ket{0}$. It is straightforward to notice,
\begin{equation}
	_r\sbra{1;m^{(1)}} \hat{H}\ket{0} = -2t \delta_{0,m^{(1)}},
\end{equation}
which means only states with $m^{(1)}=0$ angular momentum are 1-reachable from $\ket{0}$. 

In general we can introduce rotational states for higher number of moves ($n>1$) as well,
\begin{widetext}
\begin{equation}
\begin{aligned}
	\sket{n;m^{(1)},m^{(2)},...,m^{(n)}}_r = 
	\frac{1}{\sqrt{4 \cdot 3^{n-1}}} \sum_{d_1=0}^3 \exp\left(\frac{2 \pi i}{4}d_1 m^{(1)}\right) \prod_{k=2}^n \left( \sum_{d_k=0}^2  \exp\left(\frac{2 \pi i}{3}d_k m^{(k)}\right) \right) \ket{n;d_1,...,d_n}.
\end{aligned}
\end{equation}
\end{widetext}
It is easy to check that any two rotational states are orthonormal and thus they form an orthonormal basis. The two bases (position and rotational) are equivalent, i.e. one can explicitly provide the matrix that describes the change between the position basis and the rotational basis.
Note, however, that the position basis
is constructed in such a way that the created magnons are solely there due to the moving hole. Thus, such a position basis, and conseqently also the equivalent rotational basis, is only a subset of the Hilbert space of a single hole in the antiferromagnet.

\subsection{Origin of 
eigenenergies $\propto J/t$ \\
in the rotational spectra
}
\label{sec:discussionB}

The noticeable fact of the lowest rotational excitations (i.e. for which not all the $m^{(k)}$ coefficients are equal to 0) is their approximately linear in $J/t$ splitting from the ground state. Moreover, this linear splitting seems to be related to the cost of consecutive magnons created by the moving hole. In fact, in the $t$--$J^z$ model considered here (which omits the magnon-magnon interactions, see above), each magnon costs the energy of $2J$. At the same time in Figs.~\ref{fig:rot_1_no_mag}, \ref{fig:rot_11_no_mag} we can observe linear states with scaling close to $2J$ and $4J$ -- which corresponds to the cost of one and two magnons respectively.

\begin{figure}[t!]
	\begin{center}
		\includegraphics[width=\columnwidth]
		{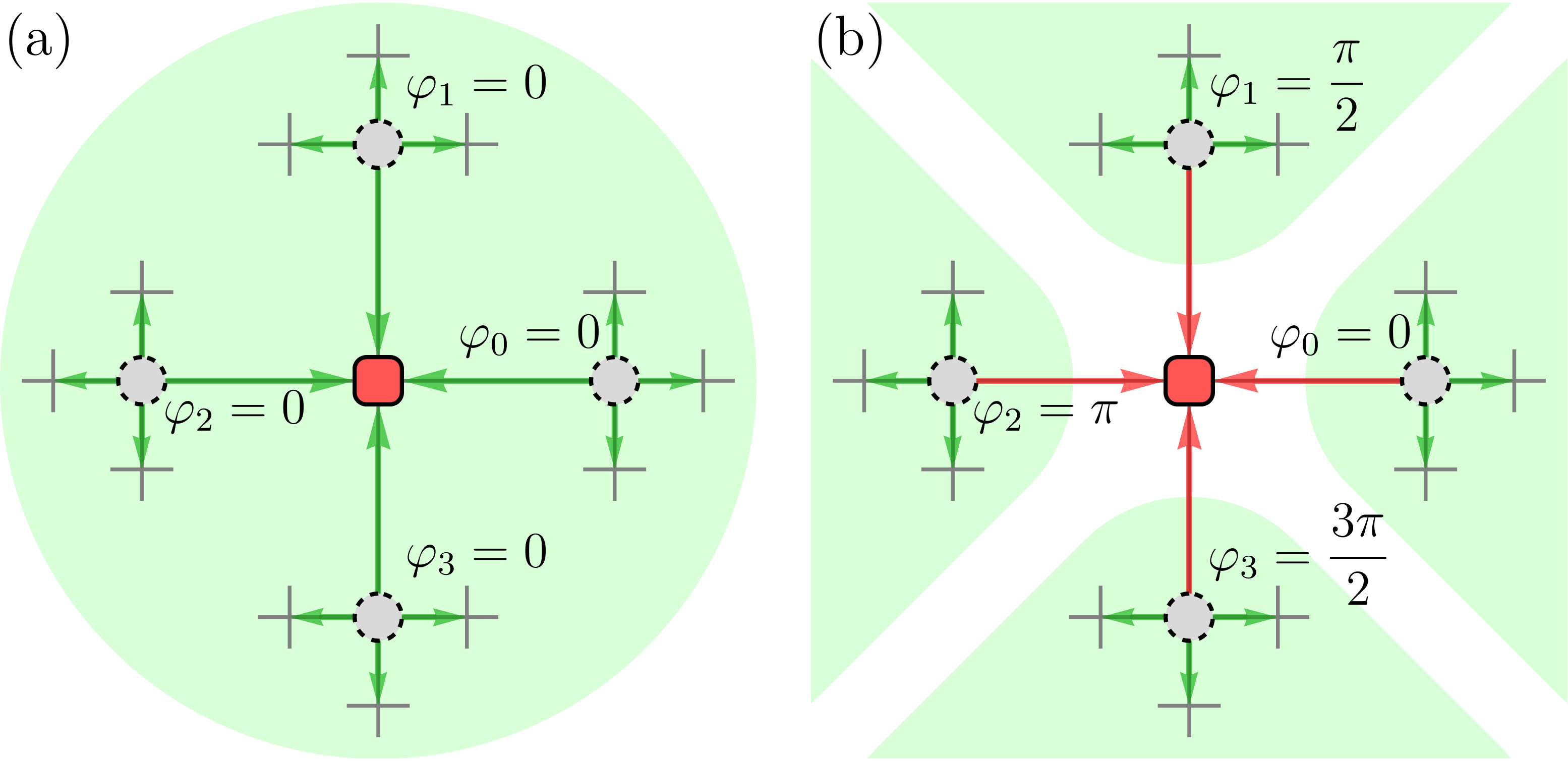}
		\caption{Cartoon picture of a single hole moving on the Bethe lattice in the \tjzm{} (in the holon-magnon basis). The hole (gray circle) is initially introduced into the vacuum state and propagated once to four nearest neighbour sites leaving a magnon (red square with rounded corners) behind: (a) without acquiring an angular momentum ($m^{(1)} = 0$), and (b) acquiring an angular momentum ($m^{(1)} = 1$). When acting on those states with the Hamiltonian the Fourier coefficients $\sum_{n=0}^3 \exp\left(i\varphi_n\right)$ sum up to (a) 4, and (b) 0, for the hole moving in the direction of the magnon. Thus the magnon cannot be annihilated in the latter case and each hole configuration is restricted to move only in a particular branch of the Bethe lattice (denoted with a green background).}\label{fig:unremovable_magnons}
	\end{center}
\end{figure}

Let us understand in quite some detail where the above-mentioned phenomenon originates from. In the beginning, let us see how the Hamiltonian $\hat{H}$ (i.e. the standard $t$--$J^z$ model) acts on the following states $\ket{n;0,...,0,0}_r$ and $\ket{n;0,...,0,1}_r$ (i.e. the latter one is the rotational state with non-zero angular momentum). To simplify equations let us consider $n>1$, but remember the same logic applies also to $n = 1$. We obtain:
\begin{equation}
\begin{aligned}
	&\hat{H}\ket{n;0,...,0,0}_r = E_n \ket{n;0,...,0,0}_r \\
	&-t\sqrt{3}\left(\ket{n-1;0,...,0}_r + \ket{n+1;0,...,0,0,0}_r\right).
\end{aligned}
\end{equation}
and
\begin{equation}
\begin{aligned}
	\hat{H}\ket{n;0,...,0,1}_r &= E_n \ket{n;0,...,0,1}_r \\ &-t\sqrt{3}\ket{n+1;0,...,0,1,0}_r.
\end{aligned}
\end{equation}
Thus we observe that once $m^{(n>1)} \neq 0$ and when $\hat{H}$ acts such that 
the hole annihilates the $n$-th magnon, the resulting 3 identical states $\ket{n-1;0,...,0}_r$ (from three paths) appear with the phase factors that sum up to 0 (cf. Fig~\ref{fig:unremovable_magnons}, where the analogous process is shown for $n=1$). Therefore there is no coupling to the states with less than $n$ magnons for $m^{(n>1)} \neq 0$. Our intuition tells us, these `unremovable' magnons shall be responsible for the observed (linear) shift in energy.

\begin{figure}[t!]
	\includegraphics[width=\columnwidth]
	{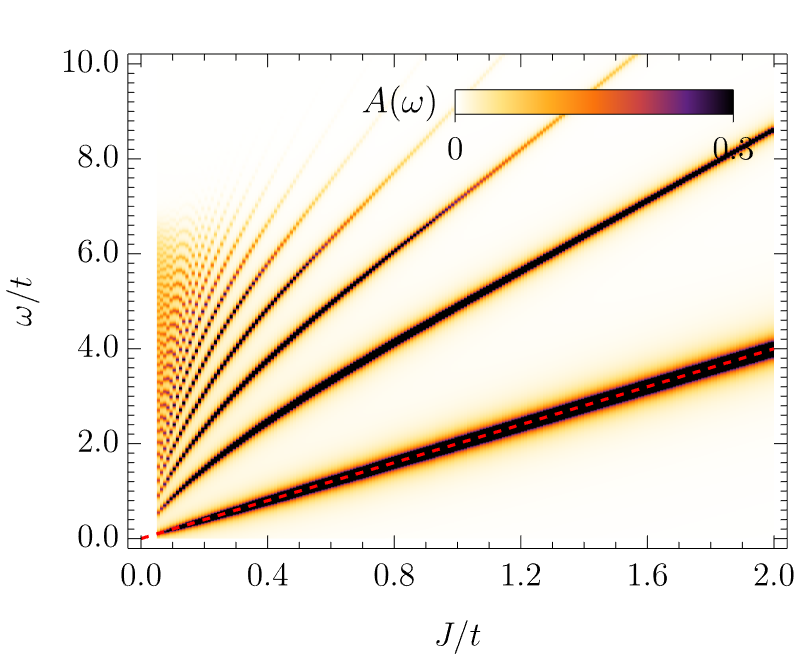}
	\caption{
		Spectral function $A_r(\omega) = -\frac{1}{\pi}\lim_{\delta \to 0^+}\mathrm{Im}(G_r(\omega + i\delta))$ aligned with respect to the ground state of the $t$--$J^z$ model with a single hole on a Bethe lattice with $z = 4$ everywhere but the single point at the origin, where $z = z_0 = 3$. The dashed red line is $\omega(J) = 2J$ and it goes through the middle of the lowest rotational excitation.
	}\label{fig:perfect_linear}
\end{figure}

\begin{figure}[t!]
	\includegraphics[width=\columnwidth]
	{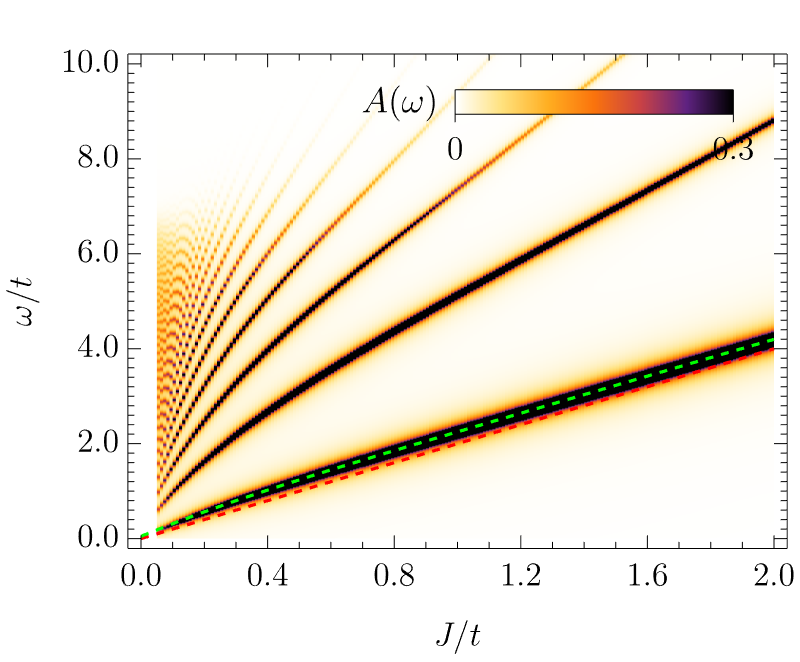}
	\caption{
		Spectral function $A_r(\omega) = -\frac{1}{\pi}\lim_{\delta \to 0^+}\mathrm{Im}(G_r(\omega + i\delta))$ aligned with respect to the ground state of the $t$--$J^z$ model with a single hole on a Bethe lattice with $z = 4$ everywhere. The dashed red line is $\omega(J) = 2J$, while the dashed green line tracks the maximum of the lowest rotational excitation: a small deviation from the linear scaling can be observed.
	}\label{fig:linear}
\end{figure}

To further confirm the above claim, let us consider the following two Greens functions,
\begin{equation}
	G(\omega) = \bra{0}(\omega - \hat{H} + E_{\varnothing})^{-1}\ket{0},
\end{equation}
which is the `standard' Greens function of a single hole added to the $t$--$J^z$~model, and
\begin{equation}
	G_r(\omega) =~_r\bra{1;1}(\omega - \hat{H} + E_{\varnothing})^{-1}\ket{1;1}_r
\end{equation}
which is the simplest spectral function including the non-zero rotational degree of freedom. We can provide the exact expressions for both of them (again for simplicity let us exclude interactions between magnons),
\begin{equation}
	G(\omega)^{-1} = G_0(\omega)^{-1} - \Sigma(\omega),
\end{equation}
where $G_0(\omega)^{-1} = \omega - 2J$ and,
\begin{equation}
	\Sigma(\omega) = \frac{4t^2}{\omega - 4J - \frac{3}{4}\Sigma(\omega - 2J)}.
\end{equation}
At the same time,
\begin{equation}
\begin{aligned}
	G_r(\omega)^{-1} 
	&= G_0(\omega - 2J)^{-1} - \frac{3}{4}\Sigma(\omega - 2J) \\
	&= G(\omega - 2J)^{-1} + \frac{1}{4}\Sigma(\omega - 2J).
\end{aligned}
\end{equation}
Notice how equations for $G(\omega)$ and $G_r(\omega)$ differ by a shift by $2J$ and some fraction of the self-energy. If we modified the geometry of the Bethe lattice only around the origin such that at this single point the coordination would be $z_0 = 3$, then we would obtain,
\begin{equation}
	G_r(\omega)^{-1} = G(\omega - 2J)^{-1}.	
\end{equation}
Therefore $G_r(\omega)$ would have the same peaks as $G(\omega)$ but shifted by $2J$ -- which is the cost of a single magnon that cannot be annihilated by the hole. This situation is presented in Fig.~\ref{fig:perfect_linear}. In detail, the reality is a little bit more complicated. When we consider the actual Bethe lattice, i.e. $z_0 = z = 4$, at the origin of the Bethe lattice the hole has 4 branches (instead of 3) via which it can delocalize and lower its energy. This pushes the ground state down in energy compared to the $z_0=3$ case. For this reason, the shift is not perfectly equal to $2J$. Moreover, the dependence of the `linear' peak is also not completely linear in $J$. But the differences are small compared to the model parameters, as we demonstrate in Fig.~\ref{fig:linear}. 

The above discussion shows how one eigenstate that scales linearly with $J/t$ can appear in the rotational spectral function. But same scheme can be generalized to higher rotational degrees of freedom. This explains the onset of several states in the rotational $t$--$J^z$ model spectra whose energy shows close-to-linear dependence in $J/t$ -- they all scale with the cost of the chain of magnons unremovable by the hole due to the non-zero angular momenta.

\subsection{Origin of 
eigenergies $\propto J/t$
in the `standard' spectra on the square lattice
}
\label{sec:discussionC}

\subsubsection{Coupling to rotational states in a toy-model}

In what follows we wish to explain how 
a single hole with zero angular momentum can couple to rotational states when moving on a square lattice. To this end we design a toy-model on a Bethe lattice that actually can mimic this phenomenon nicely -- but is far easier to grasp than the full $t$--$J^z$ model on the square lattice.

We begin by noting that for the $t$--$J^z$ model on the Bethe lattice the only $n$-reachable rotational state from the initial state $\ket{0}$ is $\ket{n;0,0,...,0}_r$. For this reason, the rotational states with non-zero angular momenta cannot be observed in the $t$--$J^z$~model (zeroth-generation) spectral function $	A(\omega)$ of the single hole on the Bethe lattice, see Fig.~\ref{fig:no_rot_no_mag}. We can modify the $t$--$J^z$ Hamiltonian $\hat{H}$ to artificially include coupling to states with non-zero angular momentum by adding a term $\hat{H}'$. The form of $\hat{H}'$ is chosen in such a way to mimic on the Bethe lattice a symmetry of some particular processes that are present for the $t$--$J^z$ model on the square lattice. In particular, the motion of the hole in a loop (or along tangential path) on the square lattice may break locally a $C_3$ symmetry -- i.e. the symmetry between three possible (equivalent) paths the hole could take to propagate (without returning). Such processes cannot occur on the Bethe lattice naturally. Our intent is not to add loops into the Bethe lattice, but rather to break the same kind of symmetry also on the Bethe lattice without having to include them.

\begin{figure}[t!]
	\begin{center}
		\includegraphics[width=\columnwidth]
		{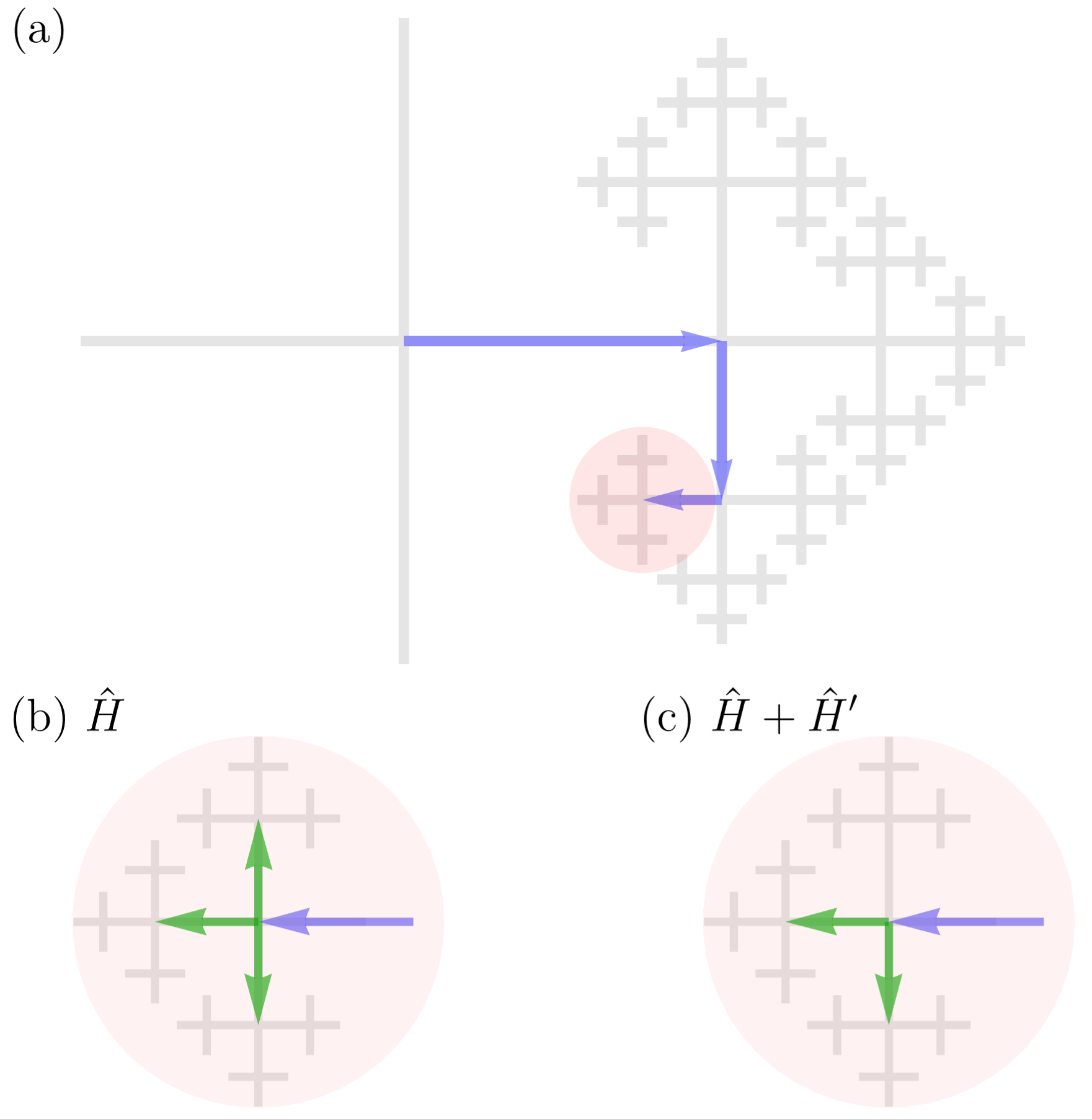}
		\caption{Cartoon picture of a hole moving in one of the branches of the Bethe lattice. Details of other branches are not included for better readability. Example path is shown in panel (a) where the hole moves in directions $E, S, W$ respectively. The state obtained by acting with $\hat{H}$ (i.e. pure $t$--$J^z$ model on a Bethe lattice)  has a local $C_3$ symmetry at the position of the hole -- as shown in panel (b) the hole can propagate forward in one of three equivalent directions. (c) This local $C_3$ symmetry is broken when $\hat{H}'$ is included -- one of three paths is blocked for the motion of the hole.
        }\label{fig:toy_model}
	\end{center}
\end{figure}

To this end, in the toy-model we introduce below we exclude 8 paths from the Bethe lattice such that the symmetry is broken the same way as due to the smallest Trugman loops on the square lattice. We do it by adding to the Hamiltonian the hopping term $\hat{H}'$, which will cancel out with the already present terms in the original model. This blocks certain hole paths and consequently breaks the $C_3$ symmetry (see Fig.~\ref{fig:toy_model}). 
Explicitly,  $\hat{H}'$ written in the position basis reads,
\begin{equation}
\begin{aligned}
\hat{H}' = t \sum_{d_1 = 0}^3 \sum_{l=1}^2 
( &\ket{3;d_1,l,l}\bra{4;d_1,l,l,l} + \\ 
  &\ket{4;d_1,l,l,l}\bra{3;d_1,l,l} ).
\end{aligned}
\end{equation}
Let us remind that $\hat{H}$ on a Bethe lattice couples only rotational states with 0 angular momenta -- e.g. consider state $\ket{3;0,0,0}_r$ which is 3-reachable from $\ket{0}$,
\begin{equation}
\begin{aligned}
	\hat{H}\ket{3;0,0,0}_r &= E_3\ket{3;0,0,0}_r \\
 &-t\sqrt{3} \ket{2;0,0}_r -t\sqrt{3}\ket{4;0,0,0,0}_r.
 \end{aligned}
\end{equation}

Now, let us observe that the perturbation $\hat{H}'$ couples to the states with non-zero angular momentum, 
\begin{eqnarray}
	\hat{H}'\ket{3;0,0,0}_r 
	&=& t \sum_{d_1 = 0}^3 \sum_{l=1}^2 \ket{4;d_1,l,l,l}\mean{3;d_1,l,l \vert 3;0,0,0}_r \nonumber \\
	&=& t \sum_{d_1 = 0}^3 \sum_{l=1}^2 \ket{4;d_1,l,l,l} \frac{1}{\sqrt{4 \cdot 3^2}}. 
	\label{eq:coupling}
\end{eqnarray}
Writing the state $\ket{4;d_1,l,l,l}$ in the rotational basis and performing possible sums we obtain,
\begin{equation}
\begin{aligned}
    t\sum_{d_1=0}^3 \sum_{l=1}^2 &\ket{4;d_1,l,l,l} \frac{1}{\sqrt{4 \cdot 3^2}} = \\
	&\frac{t}{9\sqrt{3}} \sum_{l=1}^2 \prod_{k=2}^4 \left(\sum_{m^{(k)}=0}^2  \exp\left(-\frac{2 \pi i}{3}l m^{(k)}\right) \right) \times \\
 &\times \sket{4;0,m^{(2)},m^{(3)},m^{(4)}}_r.
\end{aligned}
\end{equation}
While only states with $m^{(1)} = 0$ contribute (which is reflected in the fact, that the considered sum of the 8 states is invariant to the rotation around the origin), $m^{(k>1)}$ can have values other than 0. So the considered $\hat{H}'$ indeed couples to states with non-zero angular momentum.

\subsubsection{
Spectra of the toy-model
}

\begin{figure}[t!]
	\includegraphics[width=\columnwidth]
	{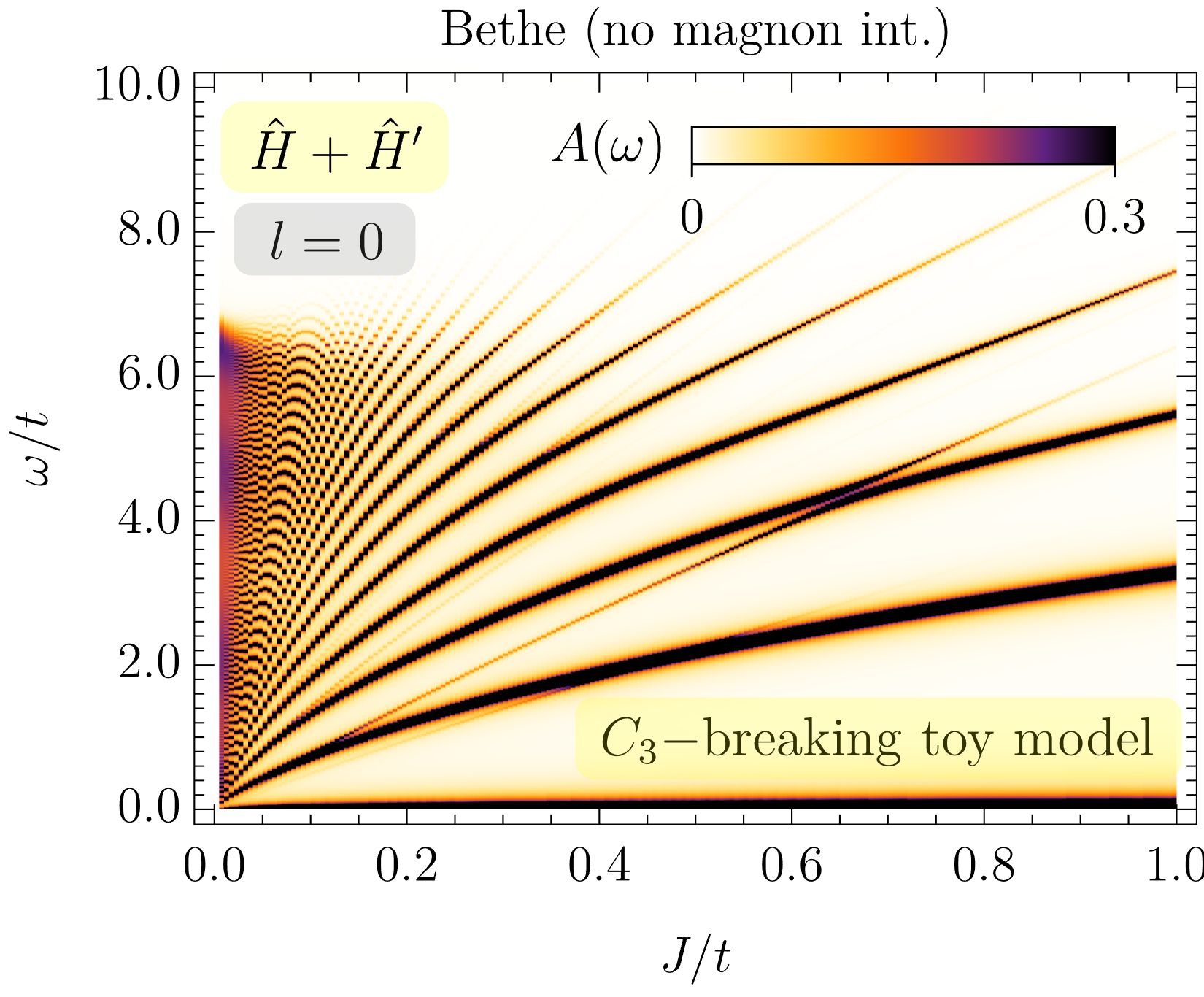}
	\caption{
		Single hole in a Bethe lattice Ising antiferromagnet: The spectral function $A(\omega)$ for the toy-model (modified $t$--$J^z$ model, $\hat{H} + \hat{H}'$, see text for further details). The coupling to states with non-zero angular momentum is reflected in the mixing of rotationally trivial and non-trivial states appearing as the level repulsion visible in the plot.
	}\label{fig:crossing}
\end{figure}

An important feature of the toy-model 
$\hat{H}+\hat{H}'$ that is defined on the Bethe lattice is the fact, that one can calculate its single-hole spectral function exactly. The result is shown in Fig.~\ref{fig:crossing}. Unlike the `standard case' without $\hat{H}'$ (i.e. $\hat{H}$ on the Bethe lattice, see Fig.~\ref{fig:no_rot_no_mag}), we observe states whose energies scale as $\propto J/t$ w.r.t. the ground state energy. In particular, one can see very clearly the linear state whose energy scales roughly like $6J$ -- this corresponds to the energy of three magnons. This state becomes visible due to the coupling and mixing of states with zero- and non-zero angular momenta -- which eventually results in observed energy level repulsion. Moreover, one can see a much weaker coupling to states that scale like $~4J/t$ and $~10J/t$. This can be understood by taking into account that states of the  $\ket{ESW}$ variety (see Fig.~\ref{fig:toy_model}) will appear with non-zero coefficient in many eigenstates of the unperturbed Hamiltonian $\hat{H}$ -- thus to some degree, all of those eigenstates will couple to the rotational eigenstates when we include perturbation $\hat{H}'$.


\subsubsection{
Remark on the $C_3$ angular momentum conservation}

Let us finish the above discussion with a remark on the conservation of $C_3$ angular momentum.
We note that the hole moving via the $t$--$J^z$ model on the Bethe lattice feels the $C_3$ symmetry on all sites, except for the origin. On the other hand, this symmetry does not hold on the square lattice {\it and} for the aforementioned toy-model on the Bethe lattice.
Thus, in the two latter cases, the angular momentum of the rotational states with nonzero $C_3$ angular momentum $m^{(k>1)}$ is not a good quantum number and does not have to be conserved.

The lack of conservation
of the $C_3$ angular momentum is in principle weakly violated. This can be inferred from the fact that the hole `predominantly' has $\tilde{z}=3$ `equivalent choices in an antiferromagnet -- and only `here and there' one of these three bonds is distinct due to the fact that the magnon had previously been created by the hole there (this leads to the violation of the $C_3$ symmetry). Interestingly: (i) this weak violation is strong enough to have a tremendous influence on the spectra -- since we observe that a {\it comparable} number of peaks in the `standard' $A(\omega)$ spectra scale linearly with $J/t$ and with $(J/t)^{2/3}$; (ii) this weak violation is small enough in order not to make the system `jump' back-and-forth between states with different $C_3$ angular momenta -- since we can nicely observe the rotational eigenstates with each particular $C_3$ angular momentum in the spectrum.

\section{\label{sec:conclusions}Conclusions}

In this work we studied the problem of a single hole added into
the Ising antiferromagnet on a square lattice. To this end, we considered
a square lattice $t$--$J^z$ model with two parameters -- $J$ being the Ising spin exchange and $t$ the hole hopping. While this is arguably the most elementary, yet non-trivial, model describing interacting spin and charge degrees of freedom beyond 1D,
its physics turned out to be far richer and complex than commonly expected.

Using self-avoiding walks approximation, that was fully benchmarked by exact diagonalization on finite clusters, 
we showed that the eigenstates of the single hole introduced into the undoped ground state of the square lattice $t$--$J^z$ model may be categorised as belonging to three distinct classes:
I. eigenstates with energies scaling as $(J/t)^{2/3}$,
II. eigenstates with energies scaling as $J/t$,
III. eigenstates with energies forming a continuum spectrum.
While class I. has been known for almost 60 years~\cite{Bul68} and class
III. was recently discussed in~\cite{Wrzosek2021,Bermes2024}, this paper showed
{\it how} class II. sets in and what are its consequences. We note in passing that the latter class was so far only believed to occur under rather special idealized conditions~\cite{Simons1990, Gru18} and its origin was not deeply understood. To make the conclusions
self-contained, below we summarize the physics behind the onset of each such a class of eigenstates: 

\begin{enumerate}[I.]
    \item We start by discussing the eigenstates with energies scaling as $(J/t)^{2/3}$.
This case is best understood by realising that we can exactly solve a closely related problem, namely that of a single hole in the $t$--$J^z$ model on a Bethe lattice (with coordination number $z=4$). As the latter one corresponds to a hole moving forward to $\tilde{z}=3$~\footnote{Except at origin site of the Bethe lattice for which $\tilde{z}=4$.}
branches of a Bethe lattice and at each step exciting a single magnon that costs energy $2J$, this problem can be exactly mapped onto a single particle moving in an external discrete linear (string) potential~\cite{Bul68, Kan89}.  
Consequently, we end up with the eigenstates that can be written in the continuum in terms of Airy functions and with the eigenergies that scale as $(J/t)^{2/3}$.
We note that in this case the hole experiences $C_3$ symmetry on all sites except for the origin. Hence, the $C_3$ angular momentum is conserved and the  $C_3$ lattice angular momenta quantum numbers $m^{(l>1)}$ 
[see Fig.~\ref{fig:rot_cartoon}(c)] can classify the eigenstates.

In the string language such `ladder' eigenstates scaling as $(J/t)^{2/3}$ are interpreted as following from the vibrational excitations of the string.

Naturally the main issue is that the motion of a hole in an antiferromagnet on the square lattice differs from the one on the $z=4$ Bethe lattice.
And it is precisely this point which leads to the onset of class II. and III. excitations. Below we explain how each of these two classes arises.

\item Let us consider that on a square lattice it {\it can} occur that, at a particular step of the hole motion, the bonds along which the hole can hop and create magnons may differ from the assumed $\tilde{z} = 3$ on the Bethe lattice.
This is due to possible creation of magnons during the `earlier' (i.e. taken during the previous steps) motion of the hole -- consequently a tangential
path may form, {\it cf}. the yellow bond of Fig.\ref{fig:saw_paths}~(b). 
Consequently, e.g. at this step the hole has only $\tilde{z}=2$ bonds along which it can move forward and create magnons. Thus, this effect {\it alters the kinetic energy experienced by the hole when moving on the square lattice w.r.t. the Bethe lattice / string potential} case.

While such a situation in principle takes place once tangential paths occur and may be considered
hard to be described, in this paper we managed to successfully explain it using the so-called rotational basis~\cite{Simons1990,Gru18}. To this end, we first realize that due to the possibility that at particular steps of the  hole motion on the square lattice $\tilde{z} \neq 3 $, 
the $ C_3$ symmetry is not conserved and the hole can move between states with different $C_3$ angular momentum quantum number
$m^{(l>1)}$. In particular, we can start with the zero $C_3$ angular momentum (as in the `standard' single-hole spectral function problem) but then the hole hops to a state with a non-zero $C_3$ angular momentum. Second, we notice that if we the hole is in a state with a finite angular momentum, then it is possible that there are `left-over' magnons which cannot be annihilated due to destructive `interference' caused by the finite lattice angular momentum and the hole cannot go back. Crucially, the latter leads to a shift of the energy of such a state by $J/t$ [note: this state also scales as $(J/t)^{2/3} $ but, due to the `unremovable' magnon, there is also an additional shift by $J/t$ w.r.t.
the ground state]. Interestingly, the violation of the $C_3$ symmetry is strong enough to make coupling to such rotational states visible in the spectra -- but small enough, in order not to couple too much to completely blur the spectra and thus make the above reasoning practically useless. That such a peculiar situation happens in such a paradigmatic many-body model, is rather anomalous.

We note that in the string language the eigenstates scaling as $J/t$ are the rotational excitations of the string.

\item Finally, let us note that it {\it can} also
occur that at a particular step of the hole motion
on the square lattice the on-site energy for the 
`just created' magnon is altered w.r.t. the one expected from the `string' potential (or Bethe lattice) analysis.
This is due to the fact that, once the hole goes along a tangential path, then the nearest neighbor magnon along the tangential path interacts with `just created' magnon and lowers its energy (alternatively, one can also understand this effect in the spin language: namely it can be accounted for by noticing that spins flipped by the hole may have lower energy if they are located next to the `already' flipped spins -- this happens along the tangential paths). 
Altogether, this effect {\it alters the potential energy felt by the hole when moving on the square lattice w.r.t. the one given by the Bethe lattice / string potential} picture. Consequently, as thoroughly discussed in~\cite{Wrzosek2021} a continuum of eigenenergies arises, {\it cf}.~\cite{Wrzosek2023} for an intuitive explanation of how flattening of a string potential may lead to the quasiparticle collapse and onset of continuum spectrum. 

This phenomenon can be understood in the string language as following from the self-interactions among strings~\cite{Bermes2024}.
\end{enumerate}

In this paper we showed that {\it all} three distinct types of states discussed above can be directly detected by single-hole spectral function and in principle measured in photoemission experiments on Ising antiferromagnets.
Furthermore, in the paper we also studied the more complex rotational spectra. These correspond to the cases
when the hole injected to the antiferromagnet is attributed with an angular momentum -- which can be of $C_4$ or the $C_3$ kind. On one hand, the obtained results turned out to be absolutely instrumental in uncovering the complex situation found in the `standard' single-hole spectra [if not for the understanding of the rotational spectra presented in Sec.~\ref{sec:discussionB}, we would not be able to explain the coupling to rotational excitations and consequently the standard spectra discussed in Sec.~\ref{sec:discussionC}]. On the other hand, these results showed how rotational spectra may give even further insight to the rich physics of the $t$--$J^z$ model, primarily by directly accessing the eigenstates with distinct (and almost conserved) angular momentum quantum numbers. 

Our work immediately raises the question whether the anomalous rotational states revealed in the spectrum of a single doped hole in an Ising antiferromagnet also exist in the canonical $t$--$J$ (or Hubbard) models, featuring ${\rm SU}(2)$ invariance, or even at finite doping. At least for the lowest-lying $C_4$ rotational excitations the answer is affirmative: Large-scale DMRG studies have recently revealed long-lived resonances in the one-rotational spectrum of a single hole in the $t$--$J$ model~\cite{Bohrdt2021PRL}. Subsequent studies have also demonstrated a rich set of rotational excitations of two paired holes in the $t$--$J^z$ and $t$--$J$ models~\cite{Bohrdt2023,Grusdt2023}, and closely related features in the spectra of Hubbard-Mott excitons~\cite{bohrdt2024arXiv}. 

Numerically, computing spectra in these more complex systems is sufficiently challenging that a study of couplings between rotationally trivial and non-trivial states, manifesting in avoided crossings in the spectra, remains an open task. Furthermore, the stability of the anomalous, long-lived excited states uncovered in this work at finite doping remains unclear. However, it has been proposed that they may play a central role in explaining the strong $d_{x^2-y^2}$ pairing in hole-doped high-$T_c$ cuprate superconductors, through an emergent Feshbach resonance~\cite{Homeier2025}. The experimental observation of non-vanishing fifth-order string-type correlations~\cite{Bohrdt2021_01} in ultracold fermion experiments up to $\sim 20\%$ doping~\cite{chalopin2024arXiv} furthermore suggests that the rotational structure of doped holes may be robust up to the most interesting doping regime of the Hubbard model. 

Last but not least, our work establishes a direct relation of doped antiferromagnets to quantum many-body scars, which constitute a special class of eigenstates beyond the eigenstate-thermalization hypothesis (ETH) which can lead to anomalously slow thermalization in certain microscopic models~\cite{Turner2018,Serbyn2021,Moudgalya2022,Chandran2023}. This phenomenon was first discovered in Rydberg tweezer arrays realizing the so-called PXP model~\cite{Bernien2017}, and subsequently a connection to string-breaking dynamics in lattice gauge theories was established~\cite{Surace2020}. The anomalous eigenstates in class II. above also fall into this category of quantum many-body scars. As we directly reveal in the spectral function, these states are long-lived and partly  co-exist with an exponentially dense set of incoherent states (class III.). This also distinguishes anomalous states from the phenomenon of Hilbert space fragmentation, another mechanism giving rise to non-ETH eigenstates. Notably, such Hilbert space fragmentation has been shown to occur in the one-dimensional $t-J^z$ model~\cite{Rakovszky2020}, but argued to be absent in higher dimensions~\cite{Moudgalya2022a}. Instead, in this paper we showed that anomalous scar-like states have approximate rotational quantum numbers which leads to their near-perfect but not complete decoupling from other states. We traced them back to rotational excitations of the doped hole featuring a rich internal string structure, which provides a natural connection~\cite{Beran1996,Grusdt2018PRX} to mesonic excitations in (lattice) gauge theories and their associated scar states~\cite{Surace2020,Iadecola2020,Banerjee2021,Aramthottil2022,Halimeh2023Scars}. Exploring this connection constitutes an interesting future research direction.

\section*{\label{sec:acknowledgements}Acknowledgements}
 We thank Takami Tohyama for stimulating discussions.
We  kindly  acknowledge  support  by the Excellence Initiative of the University of Warsaw (`New Ideas' programme) IDUB program 501-D111-20-2004310 `Physics of the superconducting copper oxides: ``ordinary'' quasiparticles or exotic partons?'.
The calculations were performed at the ICM cluster under grant no G73-29 and G88-1166.
This project has received funding from the European Research Council (ERC) under the European Union’s Horizon 2020 research and innovation programm (Grant Agreement no 948141) — ERC Starting Grant SimUcQuam, and by the Deutsche Forschungsgemeinschaft (DFG, German Research Foundation) under Germany's Excellence Strategy -- EXC-2111 -- 390814868.
K.W. thanks National Science Center, Poland for financial support 
(grant number 2024/55/B/ST3/03144).
ED acknowledges support  from the SNSF project $200021_212899$, ETH-C-06 21-2
equilibrium Grant with project number 1-008831-001, and the Swiss State 
Secretariat for Education, Research and Innovation (contract number 
UeM019-1)

For the purpose of Open Access, the authors have applied a CC-BY public copyright licence to any
Author Accepted Manuscript (AAM) version arising from this submission.

\appendix

\section{\label{sec:appendix:greens_function}Derivation of the explicit representation of the rotational Green's functions}

Let us introduce a natural notation of the directions of the world, $\hat{E},\hat{N},\hat{W},\hat{S}$, to represent possible electron hopping processes to given site $\vec{i}$ on a lattice with coordination $z=4$. Formally, we can consider $\hat{E},\hat{N},\hat{W},\hat{S}$ to be operators, which for site $\vec{i}$ can be defined as,
\begin{eqnarray}
    \hat{E}_{\vec{i}} = \sum_{\sigma} \tilde{c}_{\vec{i},\sigma}^{\dag} \tilde{c}_{\vec{i}+\hat{x},\sigma}, \\
    \hat{N}_{\vec{i}} = \sum_{\sigma} \tilde{c}_{\vec{i},\sigma}^{\dag} \tilde{c}_{\vec{i}+\hat{y},\sigma}, \\
    \hat{W}_{\vec{i}} = \sum_{\sigma} \tilde{c}_{\vec{i},\sigma}^{\dag} \tilde{c}_{\vec{i}-\hat{x},\sigma}, \\
    \hat{S}_{\vec{i}} = \sum_{\sigma} \tilde{c}_{\vec{i},\sigma}^{\dag} \tilde{c}_{\vec{i}-\hat{y},\sigma},
\end{eqnarray}
where $\hat{x} = (1,0)$ ($\hat{y} = (0,1)$) are unit vectors in positive $X$ ($Y$) direction respectively and $\vec{i} \in \mathbb{Z}\times\mathbb{Z}$ (square lattice case). For instance, with the above one can easily define e.g. an operator that moves a hole located at site $\vec{i}$ to site $\vec{i}+3\hat{x}$ through sites $\vec{i}+\hat{x}$ and $\vec{i}+2\hat{x}$ in the given order. It simply reads $\hat{E}_{\vec{i}+2\hat{x}} \hat{E}_{\vec{i}+\hat{x}} \hat{E}_{\vec{i}}$.

For further derivations, it is convenient to define $D_{\vec{i}} = \{\hat{E}_{\vec{i}},\hat{N}_{\vec{i}},\hat{W}_{\vec{i}},\hat{S}_{\vec{i}}\}$. It might be instructive to first observe how the operators $\hat{H}_t$, $\hat{R}_{\sigma, m^{(1)}}$ and $\hat{R}_{\sigma, m^{(1)}, m^{(2)}}$ defined in the main text can be written using $\hat{\xi} \in D_{\vec{i}}$ operators. For the hopping term in the Hamiltonian it is pretty straightforward,
\begin{equation}
    \hat{H}_t = -t \sum_{\vec{i}} \left( \hat{E}_{\vec{i}} + \hat{N}_{\vec{i}} + \hat{W}_{\vec{i}} + \hat{S}_{\vec{i}} \right) = -t \sum_{\vec{i}} \sum_{\hat{\xi} \in D_{\vec{i}}} \hat{\xi},
\end{equation}
where sum runs over all sites $\vec{i}$. The definition of the operator $\hat{R}_{\sigma,m^{(1)}}(\vec{i})$ is analogical,
\begin{align}\label{eq:R_1_appendix}
\hat{R}_{\sigma,m^{(1)}}(\vec{i}) = \frac{1}{\sqrt{4}}\sum_{\hat{\xi} \in D_{\vec{i}}} e^{i m^{(1)} \varphi_{\hat{\xi}} } \hat{\xi} \tilde{c}_{\vec{i}\sigma}.
\end{align}
The only new object here seems to be $\varphi_{\hat{\xi}}$. But it corresponds to the $\varphi_{\vec{j}-\vec{i}}$ in the Eq.~\eqref{eq:R_1} and its values are defined in Fig.~\ref{fig:varphi}(a), i.e. $\varphi_{\hat{E}_{\vec{i}}} = 0$, $\varphi_{\hat{N}_{\vec{i}}} = \frac{\pi}{2}$, $\varphi_{\hat{W}_{\vec{i}}} = \pi$, $\varphi_{\hat{S}_{\vec{i}}} = \frac{3\pi}{2}$. For the last operator in question, namely $\hat{R}_{\sigma, m^{(1)}, m^{(2)}}(\vec{i})$, we notice the following,
\begin{eqnarray}
    \hat{E}_{\vec{i}}^{\dag} &=& \hat{W}_{\vec{i}+\hat{x}}, \\
    \hat{N}_{\vec{i}}^{\dag} &=& \hat{S}_{\vec{i}+\hat{y}}.
\end{eqnarray}
With the above, we define,
\begin{eqnarray}
    D_{\hat{E}_{\vec{i}}} &=& D_{\vec{i}+\hat{x}} \setminus \{\hat{E}_{\vec{i}}^{\dag}\}, \\
    D_{\hat{N}_{\vec{i}}} &=& D_{\vec{i}+\hat{y}} \setminus \{\hat{N}_{\vec{i}}^{\dag}\}, \\
    D_{\hat{W}_{\vec{i}}} &=& D_{\vec{i}-\hat{x}} \setminus \{\hat{W}_{\vec{i}}^{\dag}\}, \\
    D_{\hat{S}_{\vec{i}}} &=& D_{\vec{i}-\hat{y}} \setminus \{\hat{S}_{\vec{i}}^{\dag}\}.
\end{eqnarray}
Then we can write,
\begin{equation}\label{eq:R_2_appendix}
\begin{aligned}
    &\hat{R}_{\sigma, m^{(1)}, m^{(2)}}(\vec{i}) = \\
    &= \frac{1}{\sqrt{12}}
    \sum_{\hat{\xi}_1 \in D_{\vec{i}}} \sum_{\hat{\xi}_2 \in D_{\hat{\xi}_1}} e^{i m^{(2)} \varphi_{\hat{\xi}_2\hat{\xi}_1}} e^{i m^{(1)} \varphi_{\hat{\xi}_1}}  \hat{\xi}_2 \hat{\xi}_1 \tilde{c}_{\vec{i}\sigma}.
\end{aligned}
\end{equation}
Here the values of $\varphi_{\hat{\xi}_2 \hat{\xi}_1}$ can be inferred from Fig.~\ref{fig:varphi}(b), e.g. $\varphi_{\hat{E}_{\vec{i}+\hat{y}} \hat{N}_{\vec{i}}} = \frac{4\pi}{3}$.

It is straightforward to notice that $\hat{\xi} \in D_{\vec{i}}$ will yield non-zero if and only if it acts on a state with unoccupied site $\vec{i}$. Moreover, we solely focus on the case of a single hole limit. This allows us to remove the site index with the assumption that whenever operator $\hat{\xi} \in D=\{\hat{E},\hat{N},\hat{W},\hat{S}\}$ appears, it acts on the site occupied by the only hole in the system. This greatly simplifies the notation lifting from us the necessity to deal with inconvenient site indices. Importantly, the shape of the path is fully encoded in the order of $\hat{E},\hat{N},\hat{W},\hat{S}$ operators. For instance, operator $\hat{E}\hat{E}\hat{E}$ would move the hole from its current position in the system 3 sites away in the positive $X$ direction. Note that if there were more holes in the system, the site indices would be necessary as it would be ambiguous which hole should move.

One probably could already notice that with the 
operators defined above it is possible to construct any path that connects a site occupied by the hole with any other site in the lattice. Let e.g. $\ket{\rm I} = \tilde{c}_{\vec{i}\sigma} \ket{\rm N}$ be an initial state with the hole created in site $\vec{i}$ in the undoped N\'eel state. We introduce $\hat{\eta}_l = \hat{\xi}_l \dots \hat{\xi}_2 \hat{\xi}_1$ representing a path of length $l$ that consists of moves $\hat{\xi}_n \in D$ for $n = 1,2,3,...,l-1,l$. Consistently, we follow the convention,
\begin{equation}
    \hat{\eta}_l \ket{\rm I} = \hat{\xi}_l \hdots \hat{\xi}_2 \hat{\xi}_1 \ket{\rm I} \equiv \ket{\xi_l \hdots \xi_2 \xi_1} = \ket{\eta_l},
\end{equation}
where we assume the leftmost move inside brackets to be the `youngest' one (i.e. the most recently created one). Let us notice that in general the above notation is over-complete -- the same state of the system can be represented in multiple ways, e.g. $\ket{E} = \ket{EEW}$. Fortunately, this problem does not exist if we restrict ourselves to the self-avoiding paths, meaning that the hole cannot visit the same site in a given path more than once. We will follow this assumption in our next step. This finishes the discussion of the introduced notation. Let us use it to work out the desired Greens functions of a single hole with rotational degrees of freedom,
\begin{equation}\label{eq:greens_function_appendix}
    G_{M_l}(\omega) = 
    \bra{{\rm N}}\hat{R}^{\dag}_{\sigma,M_l}(\vec{i}) 
        \frac{1}{\omega - \hat{H} + E_0}
    \hat{R}_{\sigma,M_l}(\vec{i}) \ket{{\rm N}}.
\end{equation}

Within the self-avoiding walks approximation~\cite{Wrzosek2021} the operator $\hat{R}_{\sigma,M_l}(\vec{i})$, $M_l = m^{(1)},\hdots,m^{(l)}$, (as well as the other operators, e.g. the Hamiltonian) is restricted only to self-avoiding paths (see Fig.~\ref{fig:saw_paths}). The rotational Greens function can be therefore rewritten in the following form,
\begin{equation}
    G_{M_l}(\omega) = 
    \frac{1}{\abs{\mathcal{A}_l}}\sum_{\eta_l, \eta'_l \in \mathcal{A}_l}
	\mathcal{P}_{M_l}^{\eta'_l,\eta_l}
        G_{\eta'_l, \eta_l}(\omega),
\end{equation}
where the coefficient of the Greens function is given by,
\begin{equation}
    G_{\eta'_l, \eta_l}(\omega) = \bra{\eta'_l}\hat{G}(\omega)\ket{\eta_l},
\end{equation}
and the Greens operator $\hat{G}(\omega) = ( \omega - \hat{H} + E_0)^{-1}$. The phase contribution from the angular momentum of the hole reads,
\begin{equation}
    \mathcal{P}_{M_l}^{\eta'_l,\eta_l} = \prod_{n=1}^{l} e^{-i m^{(n)} \varphi_{\hat{\xi}'_n \hdots \hat{\xi}'_{1}}} \cdot e^{i m^{(n)} \varphi_{\hat{\xi}_n \hdots \hat{\xi}_{1}}}.
\end{equation}
Consistently with introduced notation, the two states resulting from the hole traveling the paths $\hat{\eta}_l$ and $\hat{\eta}'_l$ are $\ket{\eta_l} = \hat{\xi}_l \hdots \hat{\xi}_2 \hat{\xi}_1 \tilde{c}_{\vec{0}\sigma} \ket{\rm N}$ and $\ket{\eta'_l} = \hat{\xi}'_l \hdots \hat{\xi}'_2 \hat{\xi}'_1 \tilde{c}_{\vec{0}\sigma} \ket{\rm N}$. For $l=1$ (or $l=2$) these contain the operator part of Eq.~\eqref{eq:R_1_appendix} (or Eq.~\eqref{eq:R_2_appendix}) respectively, while the phase contributions are collected in $\mathcal{P}_{M_l}^{\eta'_l,\eta_l}$. In general, $\varphi_{\hat{\xi}_n \hdots \hat{\xi}_{1}}$ depends on the geometry of the path $\hat{\eta}_l$. The exact values follow from Fig.~\ref{fig:varphi}(a) and Fig.~\ref{fig:varphi}(b) for $n=1$ and $n=2$ respectively. The states $\ket{\eta_l}$ and $\ket{\eta'_l}$ stand for the hole created at the origin $(\vec{i} = \vec{0})$ and propagated $l$ times such that the hole does not visit the same site more than once (cf. Fig~\ref{fig:saw_paths}). We denote the set of all self-avoiding paths $\hat{\eta}_l$ of length $l$ starting at site $\vec{i} = \vec{0}$ as $\mathcal{A}_l$. Accordingly, the edge case of $l=0$ is naturally defined as $\mathcal{A}_0 = \{\hat{\eta}_0 = \hat{\xi}_0 \equiv 1\}$ and $\abs{\mathcal{A}_l}$ is the number of self-avoiding paths of length $l$ starting at site $\vec{i} = \vec{0}$. To simplify the notation, we skip the hat on top of path operators $\hat{\eta}$ if we merely use them to index different functions.

The main task is to calculate the coefficient $G_{\eta'_l, \eta_r}(\omega)$ for a given pair of self-avoiding paths $\hat{\eta}_r$ and $\hat{\eta}'_l$. Although we only need $r=l$, we present below the general case, i.e. we cover the possibility of $r \neq l$. Let us denote $\hat{H}(\omega) = \omega - \hat{H} + E_0$ such that $\hat{G}(\omega) = \hat{H}(\omega)^{-1}$. Then let us define basis $\mathcal{B}$ of all possible self-avoiding walks from site $\vec{i} = \vec{0}$, i.e. states $\ket{\eta}$ of the system fulfilling $\forall_{\ket{\eta} \in \mathcal{B}} \exists_{n \geq 0}~\hat{\eta} \in \mathcal{A}_n$. In general, we can find elements $G_{\eta'_l, \eta_r}(\omega)$ with the cofactor matrix method~\cite{gantmakher1959theory},
\begin{equation}
    \mathcal{M}(\hat{G}(\omega))_\mathcal{B}^\mathcal{B} = 
    \frac{C^{T}(\omega)}{\det \mathcal{M}(\hat{H}(\omega))_\mathcal{B}^\mathcal{B}},
\end{equation}
where matrix $C = [C_{i,j}]$, coefficients $C_{i,j} = (-1)^{i+j}M_{i,j}$ and $M_{i,j}$ is the $(i,j)$-minor of the matrix $\mathcal{M}(\hat{H}(\omega))_\mathcal{B}^\mathcal{B}$ with the $i$-th row and the $j$-th column removed. With this, the problem of obtaining $G_{\eta'_l, \eta_r}(\omega)$ is replaced by the problem of calculating the ratio of the determinants of two matrices. This, although not immediately transparent, can be done with standard methods. It is thanks to the recursive structure of the matrix in the denominator and the fact that the matrix in the numerator is almost the same as the one in the denominator missing only the column corresponding to state $\ket{\eta_r}$ and the row corresponding to the state $\ket{\eta'_l}$. We spare the reader a tedious derivation. Instead, we present only the necessary ingredients required to implement the calculations of $G_{\eta'_l, \eta_r}(\omega)$. 

Let us introduce function $\mathcal{S}$ that takes a self-avoiding path $\hat{\eta}_k$ of the hole and returns set of moves such that for $\hat{\xi}_{k+1} \in \mathcal{S}(\hat{\eta}_k)$ the path $\hat{\eta}_{k+1} = \hat{\xi}_{k+1} \hat{\eta}_k$ is also self-avoiding and it is understood as an extension of the path $\hat{\eta}_k$ by a move $\hat{\xi}_{k+1}$. Then we can write a recursive equation for self-energy $\Sigma_{\eta_k}(\omega)$ of the subsystem given by all the self-avoiding paths that start with path $\hat{\eta}_k$,
\begin{equation}\label{eq:Sigma}
    t^2\Sigma_{\eta_k}(\omega)^{-1} = G_{\eta_k}(\omega)^{-1} - \sum_{\xi \in \mathcal{S}(\eta_k)}\Sigma_{\xi \eta_k}(\omega),
\end{equation}
where
\begin{equation}
    G_{\eta_k}(\omega)^{-1} = 
    \bra{\eta_k}\hat{H}(\omega)\ket{\eta_k}.
\end{equation}
Moreover we define,
\begin{equation}\label{eq:Gamma}
    \Gamma_{\eta_k}^{\Delta}(\omega) = G_{\eta_k}(\omega)^{-1} - \sum_{\xi \in \Delta}\Sigma_{\xi \eta_k}(\omega),
\end{equation}
where $\Delta \subset D$ (i.e. $\Delta$ is any subset of $D$, including the empty set). The above defined $\Gamma_\eta^{\Delta}(\omega)$ can be numerically evaluated assuming a finite depth of the recursion in Eq.~\eqref{eq:Sigma}. One can adjust the depth of the recursion to ensure convergence, provided that the value of the coupling constant is not too small, i.e. $J \gtrsim 0.2t$ for the square lattice or $J \gtrsim 0.01t$ for the Bethe lattice (as numerically checked by us). 

In the end, the generic formula for $G_{\eta'_l, \eta_r}(\omega)$ can be expressed in the following form, 
\begin{equation}\label{eq:GFcoef}
	\begin{split}
		& G_{\eta'_l, \eta_r}(\omega) = \\
		&\bra{\xi'_l \hdots \xi'_2 \xi'_1}
		\hat{G}(\omega)
		\ket{\xi_r \hdots \xi_2 \xi_1} = (-t)^{r + l - 2c} \times \\
		&\frac{\displaystyle
		\prod_{k=0}^{c-1} \left(
			\Gamma_{\eta_k}^{\mathcal{S}(\eta_k)\setminus\{\xi_{k+1}\}} - \K_{n=k+1}^{c-1} \frac{t^2}{\Gamma_{\eta_n}^{\mathcal{S}(\eta_n)\setminus\{\xi_{n+1}\}}}
		\right)
		}{\displaystyle
		\left(
			\prod_{k=0}^{c} \Gamma_{\eta_k}^{\mathcal{S}(\eta_k)}
		\right)
		\left(
			\prod_{k=c+1}^{r} \Gamma_{\eta_k}^{\mathcal{S}(\eta_k)}
		\right)
		\left(
			\prod_{k=c+1}^{l} \Gamma_{\eta_k}^{\mathcal{S}(\eta_k)}
		\right)
		},    
	\end{split}
\end{equation}
where $c$ stands for the length of the common part of $\hat{\eta}_r$ and $\hat{\eta}'_l$, i.e. $\hat{\xi}'_j = \hat{\xi}_j$ for all $j \leq c$ and either $c = \min(l,r)$ or $\xi'_{c+1} \neq \xi_{c+1}$. The symbol $\K_{n=k+1}^{c-1}$ denotes the continued fraction. Moreover, to shorten the notation we write $\Gamma_\eta^\Delta \equiv \Gamma_\eta^\Delta(\omega)$. We present an intuitive diagrammatic representation of the above equation in Appendix~\ref{sec:appendix:diagrams}.

\section{\label{sec:appendix:diagrams}Diagrammatic representation \\ of the results of Appendix A}

To grasp the essence of Eq. \eqref{eq:GFcoef} for $G_{\eta', \eta}(\omega)$ it is instructive to express it through the diagrams. To this end, let us formulate rules for creating the diagrams in terms of the introduced notation: 
\begin{enumerate}
    \item Diagram consists of nodes connected with single lines or double lines. Nodes correspond to the states of the system, while lines represent the propagation of the hole. Single lines represent propagation distinct in the left $\bra{\eta'}$ and the right $\ket{\eta}$ states while double lines correspond to common parts of the path.
    \item Each node is labeled with its corresponding weight $\Gamma_{\eta}^{\mathcal{S}(\eta)}$ where $\hat{\eta}$ denotes a path of the hole.
    \item Single lines are denoted with $-t$ as they add a factor of $-t$ to the whole solution.
    \item Double line corresponding to the move $\hat{\xi} \in D$ and starting from the node denoted with $\Gamma_{\eta}^{\mathcal{S}(\eta)}$ is denoted with $\Gamma_{\eta}^{\mathcal{S}(\eta) \setminus \{\xi\}}$.
    \item Each line ends with an arrow denoting direction of the hole motion (i.e. the order of operators $\hat{\xi}_n \in D$) in the left and the right state.
\end{enumerate}
To read out the expression from the diagram we simply multiply in the denominator weights of all the nodes. For each single line we put in the numerator the factor of $-t$. And the most complex rule is for the double lines. For each double line, we multiply the numerator by a finite continued fraction---this continued fraction is formed from the weight of the current double line and all following double lines down the path. Let us give some examples.

With the above, the `standard' local Greens function can be written as follows,
\begin{equation}
    G_{1,1}(\omega) = \frac{1}{\Gamma_1^{D}(\omega)},
\end{equation}
or expressed through the simplest diagram.
\begin{center}
	\begin{tikzpicture}[
	roundnode/.style={circle, draw=black!60, fill=black!5, very thick, minimum size=0.25}
	]
		\node[roundnode]    
			(m0)                    {$\Gamma_1^{D}$};
			
	\end{tikzpicture} 
\end{center}
For the coefficient $G_{EE,EE}$ the diagram is as follows,
\begin{center}
    \begin{tikzpicture}[
    roundnode/.style={circle, draw=black!60, fill=black!5, very thick, minimum size=0.25}
    ]
        \node[roundnode]    
            (m0)                    {$\Gamma_1^{D}$};
        \node[roundnode]    
            (mE)    [right=of m0]   {$\Gamma_E^{D \setminus \{W\}}$};
        \node[roundnode]    
            (mEE)    [right=of mE]   {$\Gamma_{EE}^{D \setminus \{W\}}$};
        
        \draw[->, >=stealth, draw=black!60, very thick, double] 
            (m0.east) -- node[above] {$\Gamma_1^{D \setminus \{E\}}$} (mE.west);
        \draw[->, >=stealth, draw=black!60, very thick, double] 
            (mE.east) -- node[above] {$\Gamma_E^{\{N,S\}}$} (mEE.west);
    \end{tikzpicture}
\end{center}
so we can read out the expression,
\begin{equation}
	G_{EE,EE}(\omega) = \frac{
	\left(
	\Gamma_1^{D \setminus \{E\}}
	-\frac{t^2}{\Gamma_E^{\{N,S\}}}
	\right)
	\Gamma_E^{\{N,S\}}
	}
	{
		\Gamma_1^{D}
		\Gamma_E^{D \setminus \{W\}}
		\Gamma_{EE}^{D \setminus \{W\}}
	}.
\end{equation}
Similarly for $G_{NE,EE}$,
\begin{center}
    \begin{tikzpicture}[
    roundnode/.style={circle, draw=black!60, fill=black!5, very thick, minimum size=0.25}
    ]
        \node[roundnode]    
            (m0)                    {$\Gamma_1^{D}$};
        \node[roundnode]    
            (mE)    [right=of m0]   {$\Gamma_E^{D \setminus \{W\}}$};
        \node[roundnode]    
            (mEE)    [right=of mE]   {$\Gamma_{EE}^{D \setminus \{W\}}$};
        \node[roundnode]    
            (mEN)    [below=of mEE]   {$\Gamma_{NE}^{D \setminus \{S\}}$};
        
        \draw[->, >=stealth, draw=black!60, very thick, double] 
            (m0.east) -- node[above] {$\Gamma_1^{D \setminus \{E\}}$} (mE.west);
        \draw[->, >=stealth, draw=black!60, very thick] 
            (mE.east) -- node[above] {$-t$} (mEE.west);
        \draw[->, >=stealth, draw=black!60, very thick] 
            (mE.south) -- node[above] {$-t$} (mEN.west);
    \end{tikzpicture}
\end{center}
we can read out the formula from its diagram,
\begin{equation}
	G_{EE,NE}(\omega) = \frac{
	t^2\Gamma_1^{D \setminus \{E\}}
	}
	{
		\Gamma_1^{D}
		\Gamma_E^{D \setminus \{W\}}
		\Gamma_{EE}^{D \setminus \{W\}}
		\Gamma_{NE}^{D \setminus \{S\}}
	}.
\end{equation}
The above expressions should be compared with Eq.~\eqref{eq:GFcoef}.

Moreover, while performing the calculations one shall take the advantage of the symmetries of the system, e.g. $G_{EE,EE} = G_{NN,NN} = G_{WW,WW} = G_{SS,SS}$, to reduce the number of terms that have to be calculated to find the Greens function with rotations.

\section{\label{sec:appendix:ED}
Benchmarking the SAW approximation using exact diagonalization}

\begin{figure}[t!]
\begin{center}
    \begin{minipage}[c]{\columnwidth}
        standard spectral function\\
        w/o magnon-magnon interactions\\
    	\includegraphics[width=0.8\columnwidth]
	{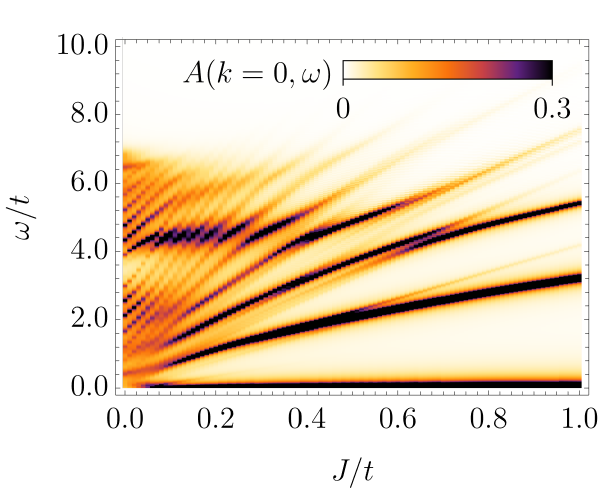}
    \end{minipage}
\end{center}
\caption{Dependence of the (local) spectral function $A(\omega )$ of a single hole in the Ising antiferromagnet on the coupling constant $J/t$. Results obtained {\it without} magnon-magnon interactions and on the square lattice with 20 sites using exact diagonalization.}\label{fig:ED_20nmm}
\end{figure}

\begin{figure}[t!]
\begin{center}
    \begin{minipage}[c]{\columnwidth}
        standard spectral function\\
        w/ magnon-magnon interactions\\
    	\includegraphics[width=0.8\columnwidth]
	{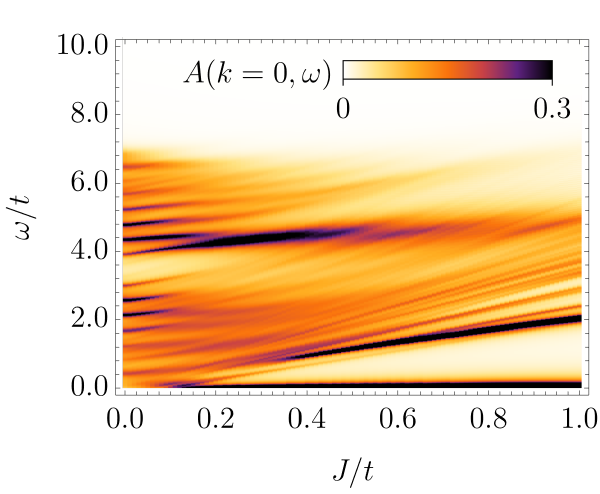}
    \end{minipage}
\end{center}
\caption{Dependence of the (local) spectral function $A(\omega )$ of a single hole in the Ising antiferromagnet on the coupling constant $J/t$. Results obtained after including magnon-magnon interactions and on the square lattice with 20 sites using exact diagonalization.}\label{fig:ED_20}
\end{figure}

To confirm that the class of states linear in $J/t$ is not merely the side effect of the SAW approximation  we perform exact diagonalization of the \tjzm{} on the square lattice with 20 sites in the unit cell. Our results without (with) magnon-magnon interactions are shown in Fig.~\ref{fig:ED_20nmm} (Fig.~\ref{fig:ED_20}). Despite larger broadening ($\delta = 0.05t$) compared to the figures in the main text ($\delta = 0.01t$) we can clearly see features that scale linearly in $J/t$. Thus, crucially the results are on the qualitative level similar to Fig.~\ref{fig:no_rot_no_mag}a (Fig.~\ref{fig:no_rot}a). But there are some differences that originate from either a finite size effect or the existence of Trugman loops. The most visible difference can be seen along all the $J/t$ values at energy slightly larger than $4t$ above the ground state. We can see there a largely pronounced spectral weight in the spectrum. In the same time it looks like peaks fall down to lower energy states at the weight bumps. Moreover, we can see the same feature repeats in higher energies although with smaller amplitude.

%

\end{document}